# Turbulence characteristics of wave-blocking phenomena

**Debasmita Chatterjee, B. S. Mazumder* and Subir Ghosh**

Fluvial mechanics Laboratory, Physics and Applied Mathematics Unit
Indian Statistical Institute, Kolkata 700108, India

*Corresponding author: bijoy@isical.ac.in Ph. No. +91-9831175336

## Abstract

This study explores experimentally the turbulent flow in a laboratory flume, interacting with waves propagated against the flow. It focuses a region of 'wave-blocking' for which there is a stream-wise location on the water surface, where the wave propagation velocity vanishes. The observations are corroborated by finding a critical wave frequency for a particular discharge above which the waves are effectively blocked; and verified by the dispersion relation of monochromatic wave. The counter-current propagating waves show an evolutionary change in the flow with three segmented regions, viz, flow at the upstream, blocking at the mid-stream and waves in the downstream. The instantaneous velocity data were collected using 3-D Micro-acoustic Doppler velocimeter (ADV) along the flume centerline. This study addresses the changes in the mean flows, Reynolds stresses, eddy viscosity, turbulence kinetic energy fluxes and associated contributions of burst-sweep cycles to the total Reynolds shear stress due to addition of surface waves against a current. The velocity power spectral analysis shows the energy distribution over the whole profile from upstream to downstream. The quadrant analysis is also used to highlight the turbulent event evolutions along the flow; and shows that at the wave-blocking and wave dominated regions, the contributions from ejection and sweep to the total shear stress are dominant. The changes in turbulence key parameters due to wave-blocking may affect the sediment transport in coastal region.

## HIGHLIGHTS

1. Study of 'wave-blocking' for waves propagating against a current.
2. Non-uniformity of flow with wave-blocking and wave dominated regions.
3. Turbulence properties of counter-current propagating waves.
4. Mean flows, Reynolds stresses, spectra and conditional statistics are analyzed.

# 1. Introduction

Fluid flows observed in coastal environments are usually a combination of waves and currents. The combined wave-current flows govern many physical processes of interest to oceanographers and engineers working in coastal environments. For example, at the mouth of a tidal inlet the interaction between waves and tidal current can cause harms in navigation, affect the design of coastal structures, and alter the transport of sediment in the near shore region. Unsurprisingly, this interaction is of significance to coastal engineering and oceanography. As the waves attempt to penetrate into an opposing current, its group velocity reduces. This causes an increase in wave height and a subsequent decrease in wavelength that can induce the waves to break causing energy dissipation. In the process wave field can be blocked if its group velocity goes to zero for sufficiently strong opposing current. Thus, to understand the physical processes arising from wave-current interacting flows, several investigations have been carried out over the last few decades for waves following or opposing a current both theoretically and experimentally (e.g. Van Hoften and Karaki [1], Grant and Madsen [2], Brevik and Aas [3], Kemp and Simons [4, 5], Klopman [6], Mathisen and Madsen [7, 8], Umeyama [9, 10], Mazumder and Ojha [11], Ojha and Mazumder [12], Singh et al. [13].

For example, Brevik and Aas [3] presented the mean flow and friction factor arising from both following and opposing currents with surface waves over ripple beds. An experimental study for wave-current flows over smooth and rough boundaries was reported by Kemp and Simons [4], where waves propagated along the current. In their subsequent study [5], they focused the results, when the waves propagated against a current over the same boundaries, and compared the results with that of waves following a current. They reported that the mean velocity near the bed was insensitive to whether the wave propagated following or against current. For waves along the current, an increase in the bed shear stress was observed at the rough boundary, while for an opposing current, the near-bed velocities for the rough surface were decreased by ~ 40%. The near-bed turbulence intensities for both rough and smooth boundaries were increased by 50%. In combined wave-current flows, Van Rijn et al. [14] reported that the presence of irregular waves causes a decrease in mean velocity in the near bed region, but waves following current decreases the near surface velocity and vice-versa. The similar results were also obtained by Klopman [6] on wave-current interactions. Two-dimensional wave-current interaction model applicable to the weak current was studied by Nielson and You [15] using a stress balance concept. The combined wave-current flows over the ripple bed were investigated experimentally by Mathisen and Madsen [7, 8] to determine the bottom roughness for combined boundary layers. They showed that the apparent hydraulic roughness was underestimated as obtained by Grant and Madsen [2]. Umeyama [9, 10] performed a series of experiments in an open channel flow for only current, waves following and opposing a current; and analyzed the changes of mean velocity and turbulence. They reported that for waves opposing current, wave attenuation increases, resulting loss of wave energy, and hence increase in mean flow near the water surface. On the other hand, for waves following current, there is less wave attenuation, less loss of energy and reduction in mean flow near the water surface.

Based on previous studies, it is assured that several experiments on wave-current interactions with waves following or opposing current had been performed over the years. A few attempts have been made to study for waves propagating against a current with 'wave-blocking' condition theoretically as well as experimentally, for which there is a stream-wise location where the wave propagation velocity vanishes. For example, a river mouth that empties in a sea, the river flow blocks the incoming sea waves. Chawla and Kirby [16] performed a series of experiments to understand dynamics involved in the interaction between waves and strong opposing currents. They reported some interesting observations due to the interaction between waves and currents including frequency downshift and complete blockage of waves. It may be mentioned here that for sufficiently strong opposing current with the waves, the group velocity goes to zero, causing the waves to be blocked. They also focused their study to the energy dissipation due to monochromatic and random wave breaking at or before blocking points, and proposed a dissipation relation. Ma et al. [17] carried out experiments in a laboratory flume to study the nonlinear evolution of regular waves with different initial periods and different wave steepness $s$ ($0.05 < s < 0.19$) propagating on spatially varying opposing current along the flume. In the complete blockage of waves due to opposing current, they observed the peak-frequency downshift even in waves with small initial steepness ($s < 0.10$), which was consistent with results of Chawla and Kirby [16]. Recently, Shugan et al. [18]

developed a weakly nonlinear analytical model of Stokes waves on non-uniform unidirectional current. They focused their results to the stationary non-dissipative solutions for adverse and following non-uniform currents with different wave steepness. The interaction of steep surface waves with the strong adverse current under the wave-blocking conditions including wave breaking was also presented and compared their results with [16, 17]. Haller and Ozkan-Haller [19] simulated the effects of current-induced breaking on the modelling of wave breaking dissipation that occurs due to rib-currents; and these are compared with laboratory data. Recently, Soltanpour et al. [20] conducted experiments to study the attenuation of waves with current over the mud layers, and studied the wave spectra for all three cases.

The knowledge of turbulence properties associated with interactions of surface waves with opposing current generating wave-blocking condition is important for coastal erosion, navigational hazard, river planning and restoration, bed form migration near inlet/river-mouth and channel evolution process. The results arising from the 'wave-blocking' in the present study may imitate the blocking phenomenon generated in coastal environments, where the river flow blocks the incoming sea waves at the river mouth. Therefore, a substantial investigation is required to examine the basic single-phase turbulence and the flow characteristics in the wave-blocking. This investigation has the potential to be useful to the researchers who study the turbulence in wave-current environments in coastal regions.

The objective of this study is to investigate experimentally in a flume the mean flow and turbulence properties in wave-current interacting flows, when the waves propagate against a current at a wave-blocking condition. The counter-current propagating waves show an evolution of flow pattern with three segmented flow regions, such as, flow at the upstream, the wave-blocking at the mid-stream and the wave region at the downstream. More precisely, an attempt has been made to detect a wave-blocking region from a certain pair of flow discharge and frequency of opposing wave; and to address the associated turbulence parameters, such as, mean flows, Reynolds stresses, turbulence spectra, eddy viscosity, turbulence kinetic energy, and the fractional contributions of burst-sweep cycles to the total shear stress along the three regimes of flow. The novelty of this study is to explore the turbulence statistics of basic flow and wave-current interacting flow under the complex wave-blocking environment; and their comparative study. Therefore, it is important to study in detail the turbulence phenomena from the fluid velocity spectrum under combined flows, especially for wave-blocking condition. The possible effects of wave-current interaction on sediment transport were suggested by Kemp and Simons [4] for current following waves, Ribberink [21] for both following and opposing waves, and Soltanpour et al. [20] for all cases in the flow. The knowledge of flow against the surface wave still has significant shortcomings, because there are lots of unknown physics of flow subjected to the frontal collision with the counter propagating waves, which are yet to be studied.

The descriptions of experimental set up for only current and wave-blocking are provided in Section 2; data processing and analysis is presented in Section 3; Dispersion relation for wave-blocking and verification with observed data in Section 4; Results and discussions including mean flows, turbulence intensities, Reynolds shear stress, spectral analysis, TKE fluxes, eddy viscosity, and coherent structures are presented in Section 5; and Summary and conclusions in Section 6.

## 2. Experimental set up and measurements

*2.1 Test Channel*

Experiments were conducted in a specially designed re-circulating flume (Mazumder et al. [22], Sarkar et al. [23]) at the Fluvial Mechanics Laboratory, Physics and Applied Mathematics Unit, Indian Statistical Institute, Kolkata. The sidewalls of the experimental flume were made of Perspex windows with a length of 8.5 m, providing a clear view of the flow. The flume consisted of both the experimental and re-circulating channels of same dimensions (10 m long, 0.5 m wide and 0.5 m deep), which looked like oblong in shape (top view, Fig. 1a). Here in this flume, the commonly used narrower re-circulating pipe located below the experimental channel was avoided. Water was put into the re-circulating flume at a desired depth for experiment. The main advantage of the present flume was that the whole water body was re-circulated throughout the flume without any disruption of flow due to passing through the narrower cross-sectional pipe. Two non-clogging types of centrifugal pumps for the flow discharge were located outside the main body of the flume. The intake and outlet pipes were freely suspended to allow tilting the flume. The outlet pipes (Pump-1 and 2) were fitted with by-pass pipes and valves, so that the flow discharges were adjusted. The electromagnetic discharge meters are fitted with the outlet pipes for continuous monitoring of the flow. The upstream bend of the channel is divided into three sub-channels of equal width 0.165m in dimension. Moreover, two honeycomb cages are placed at the back end of the sub-channels in front of the jets of high flow coming out from the outlets, and one honeycomb cage is placed at the other end of the sub-channels in order to ensure the vortex free and uniform flow of water through the experimental channel. The positions of cages in front of the jets and at the upstream end of the experimental channel substantiate the curvature effect free flow from the measurements.

*2.2 Wave-Maker*

A plunger-type wave-maker is mounted at the downstream end of the flume at a distance 8.5 m from the source to generate surface waves against the current (side view, Fig. 1b). Two six-inch wheels are fitted at the end of a spindle, which has got a gear in its middle position. One crank and shaft is connected at the rim end of each wheel. The shafts are allowed to pass through a guide to restrict their motion in vertical direction only. A triangular shaped six-inch cylinder closed at both ends is fitted at the other ends of the shafts, which is similar to the wave-maker of Euve et al. [24]. When the spindle is rotated using motor, the cylinder moves up and down to generate waves. The wave-maker is placed in such a way that the triangular cylinder remained partially submerged in water when it is at its extreme positions: topmost and lowermost. This is done to avoid the generation of small unwanted waves and disturbances in the flow. Oscillatory wave is generated at a right angle to the steady unidirectional current, which leads to surface waves propagating against current. The wave-maker is fixed with a Variac (variable resistor) to control the frequency of oscillation. Calibration is made using tachometer for frequency variation. As the flume used in the present study is a re-circulating, the generated surface waves may get re-circulated with the flow, which may bring more complexity in the flow. To avoid this complexity, a wave-absorber is placed further downstream behind the wave-maker. Here the digital camera is used to record the wave-lengths and amplitudes of waves travelling against the flow.

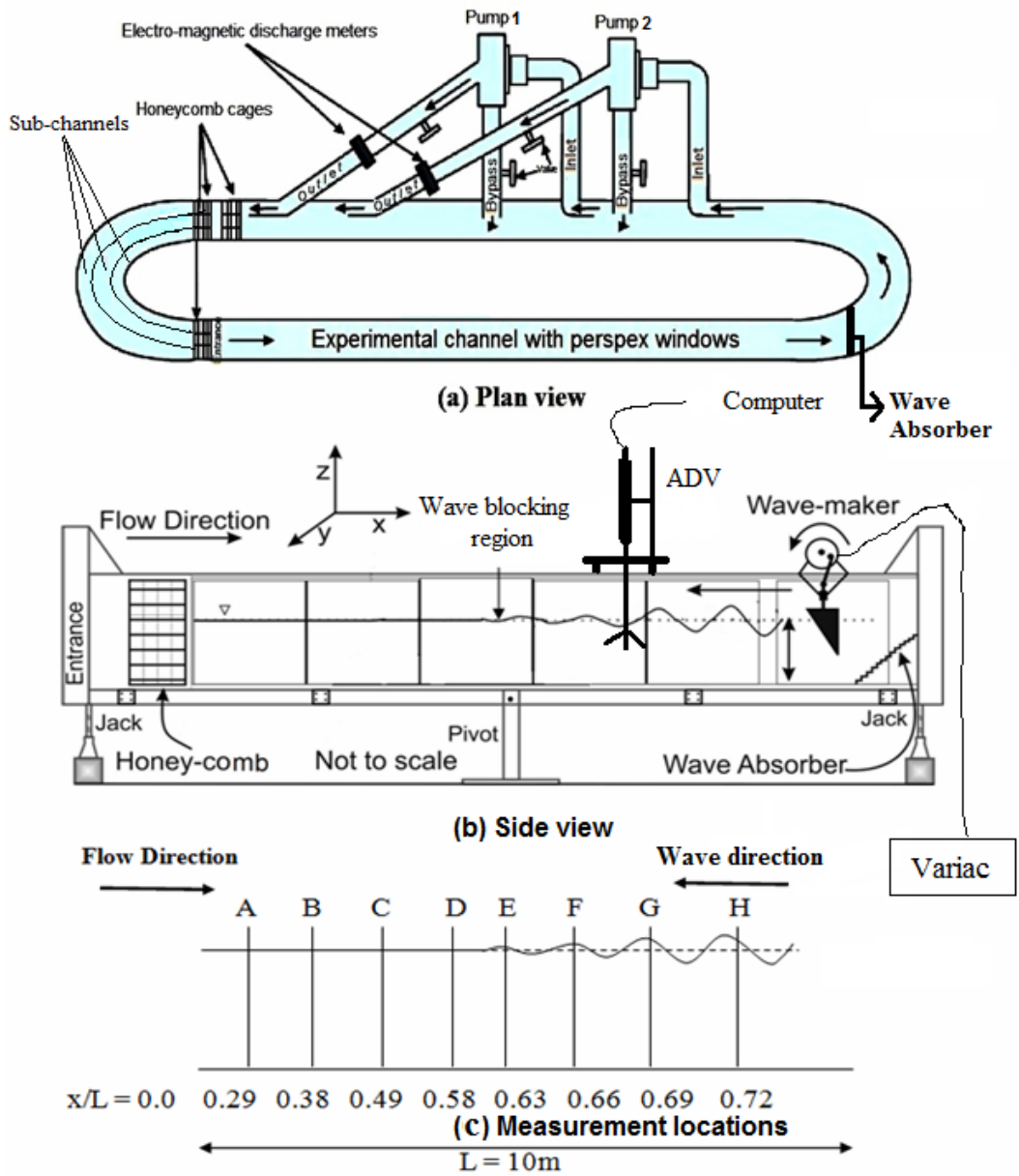

Fig. 1: Schematic diagram of the flume (a) plan view (after Fig.1 of Sarkar et al. [23]), (b) front view with wave-blocking condition, and (c) measurement locations. Here A, B, C, ......,H indicate locations of velocity data collection along the flow. Locations A, B, C denote the flow region; D, E and F denote the wave-blocking region with D as transition; and G, H denote the wave-dominated region.

*2.3 Flow measurements*

The experiments were carried out at a desired discharge in the flume in two steps: (1) to verify the fully developed flow over the rigid flat surface, measuring flow velocity at eight different vertical locations along the centerline of the channel, and (2) to measure the flow velocity in wave-blocking condition, when waves are superimposed against a current over the rigid flat surface, and to make a comparative study with the basic turbulence of fully developed flow. The coordinate system of measurement is as follows: *x* along the flow, *y* transverse to the flow towards left wall and *z* is bottom-normal to the flow positive upward. The origin (0, 0, 0) is at the inlet source at the centerline near the honeycomb.

First, experiments were conducted over the rigid flat surface at a flow discharge $Q = 0.0322 \, m^3/s$ to confirm the fully developed flow in the channel. Eight different measuring stations A, B, C, D, E, F, G, and H along the flow from upstream to downstream were selected (Fig.1c, side view); and the respective dimensionless distances of the locations were x/*L* = 0.29, 0.38, 0.49, 0.58, 0.63, 0.66, 0.69 and 0.72, with x/*L* = 0 at the inlet source, where *L* = 10 m is the length of the experimental channel. Water depth was kept constant at *h* = 30 cm for all experiments and the hydraulic slope of the flume was of order 0.0001. After a passage of certain time about 60 minutes, when the flow achieved an equilibrium state, the instantaneous velocity data were collected using a SonTek 16MHz down-looking 3-D acoustic Doppler velocimeter (ADV) from the lowest level 0.35 cm to the highest level ~24 cm for 180 seconds at a sampling rate of 40 Hz from all the eight vertical locations along the flume centreline. The sampling volume was located 5 cm below the transmitter probe and the entire probe was immerged in water. ADV did not allow collecting velocity data near the free surface. The velocity data were cleaned by removing communication errors, low signal to noise ratio (< 15 dB) and low correlation samples (< 70%). Below the level 0.35 cm, sampling height was undetected, and hence the measurements of velocity data were erroneous (SonTek Inc. [25]). The ADV sampling volume is 9 x $10^{-8}$ $m^3$ and is approximately cylindrical oriented along the transmitter beam axis. It has a diameter equal to that of 0.6 cm ceramic of the transmitter (Lohrmann et al. [26]), and the vertical length is 0.32 cm. Factory calibration of the ADV is specified to be ±1.0% of the measured velocity (i.e., an accuracy of ±1 cm/sec is on a measured velocity of 100 cm/sec).

The state of equilibrium condition of flow at a point in a location was confirmed from three repeated samplings of instantaneous velocity data for each interval of 60 sec using the ADV. The time series of instantaneous velocity data of 60 sec at a point are shown in Fig. 2 (a, b, c), and their time-averaged velocity and variance values show almost same at that point, indicating the stable state of flow. Thereafter, in the present experiments the instantaneous velocity data were continuously collected using ADV for different lengths of time period, like three minutes, four minutes and five minutes of intervals, to achieve the stable statistics of flow; and found that the time-averaged of velocity data for each period of time showed almost same values of mean and variance. Finally, the collection of velocity data was maintained for three minutes of interval to reproduce the turbulent coherent structures in these experiments, and to satisfy the ergodicity condition with reference to the velocity mean and variance (Bendat and Piersol, [27]). Therefore, for the present study, the duration of three minutes of velocity data was selected throughout the analysis of the problem. In their experiments, Barman et al. [28] followed the same procedure for collection of velocity to achieve the stable statistics of flow.

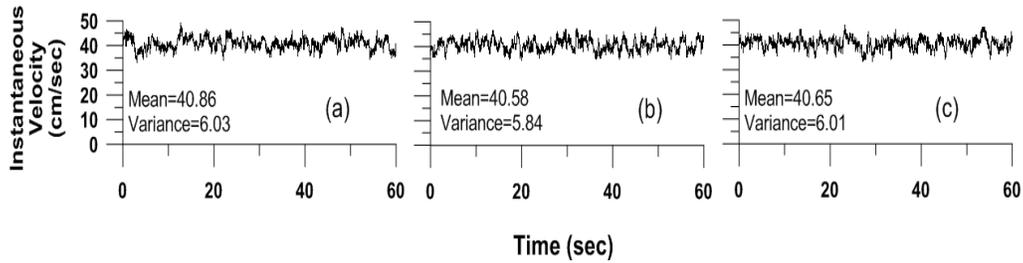

**Fig. 2** Instantaneous velocity plots of 60 sec each, showing the equilibrium state.

Once the fully developed flow was achieved, the experimental set-up was made in the flume with counter-current propagating waves. The existence of a critical frequency ($\omega_c$) of incidence wave at a particular flow discharge ($Q$) was observed for which the wavelength was effectively blocked at a certain stream-wise location. The 'wave-blocking' phenomenon emerged, when the counter-current became sufficiently strong to block the upstream propagating wave. The propagation of waves against a current was blocked at a location, where the opposing current reached the group velocity of the wave; and the wave could not penetrate over the blocking. The incoming wavelength mode was converted to two short wavelengths that were brushed away with the flow and could not travel beyond that blocking, that means, the counter-current propagating waves induced a mode conversion. The incoming incident long wavelength (that was impressed externally) interacting with a counter-current produced co- and counter propagating short wavelength modes with different wave numbers under the same frequency of stationary coordinate (Maïssa et al. [29]). Near the blocking zone one short wavelength wave was reflected back towards downstream and the other was transmitted towards upstream, indicating the energy of incoming wave was not conserved as it converted into other modes. However, it can be shown that the wave-action flux is conserved, which is the extreme importance of physical process involved in the combined wave-current flow. Essentially the shallow water wave reaches the blocking zone, gets blocked and is converted to two deep water waves which are sweep back with the background flow. For a pair of discharge ($Q$) and critical frequency ($\omega_c$) of wave, a non-uniform flow was observed along the flume with a 'flow-only' at the upstream, 'wave-blocking' at mid-stream, and 'wave-dominated region' at the downstream. The locations A, B, C indicate the flow-only region; the locations D, E, F indicate the wave-blocking; and locations G, H indicate the wave-dominated region. We were interested to address how the turbulence key parameters along the flow were modulated due to the wave-blocking over the rigid flat surface. Therefore, here experiments was conducted to analyze the turbulence in a quasi-steady state for three pairs of discharge and critical frequency ($Q$, $\omega_c$), such as ($0.0197\, m^3/s$, 1.40 Hz); ($0.0250\, m^3/s$, 1.31 Hz); and ($0.0322\, m^3/s$, 1.30 Hz), that formed a wave-blocking at a mid-stream location.

Similarly, again after a certain time about 60 minutes, when the wave-blocking flow condition along the flume achieved an equilibrium state, the instantaneous velocity data were recorded using ADV from all eight vertical locations from the lowest height 0.35 cm to the highest level ~24 cm along the flow in three respective regions for different Reynolds numbers $\mathrm{Re} = \dfrac{u_m h}{\nu}$, and the Froude numbers $Fr = \dfrac{u_m}{\sqrt{gh}}$, where $u_m$ is the maximum mean velocity, $\nu$ is the kinematic viscosity, and $g$ is the acceleration due to gravity. The equilibrium condition of flow during wave-blocking in the flume was tested from three repeated samples of instantaneous velocity data collected at locations C, D and E for each interval of 60 sec using the ADV for all three different Reynolds numbers (Re). The time series of 60 sec velocity data at locations C, D and E were analyzed, and it is observed that time-averaged velocities and variances for each time slot showed almost same, confirming the stable state of blocking condition. Here, time series plots at the location D ($x/L$ = 0.58) at three different vertical levels ($z/h$ = 0.015, 0.55, 0.75) are shown in Fig. 3. Each column in Fig. 3 represents the time slot of 60 sec. Therefore, in a similar manner as above in fully developed flow, three minutes of instantaneous velocity data collected continuously were selected to analyze for the investigation of turbulence characteristics. For photographic purpose, the transparent graph sheets having the grid size (*mm* x *mm*) were pasted in the Perspex wall of the flume from outside. The digital camera was used to record several images of amplitudes and wave-lengths of opposing wave frequency travelling over the water surface at the wave-dominated region. Four identical images with complete two wave cycles were chosen to estimate the surface wave-length ($\lambda$) and the wave-amplitude ($h_w$). The pixel units of images were converted into metric units. The values of wave-length and wave-amplitude were estimated finally by averaging the data of four identical images; and presented in the paper (Table-1). It may be noted that the calculated flow discharge ($Q_{cal}$) from the maximum mean velocity ($u_m$) and wetted perimeter is approximately 1.86 times greater than recorded flow discharge ($Q$) using electromagnetic discharge meter (Table-1). The recorded flow discharge ($Q$) is referred throughout the paper; and the results are presented only for one pair of flow discharge and frequency ($0.0322\, m^3/s$, 1.30 Hz). According to Dean and Dalrymple [30], whether the wavelength $\lambda$ is much shorter or longer than the mean water depth $h$ were computed for different flow discharges ($Q$ = 0.0197, 0.0250, $0.0322\, m^3/\text{sec}$). Table-1 shows the values of mean flow depth ($h$), wavelength ($\lambda$), initial wave-amplitude ($h_w$), critical wave frequency ($\omega_c$), wave steepness ($s$), and wave nature for three pairs of discharge ($Q$) in the opposing wave-dominated region. The incoming wave-amplitude ($h_w$) decreases as approaching towards the blocking region and diminishes to millimeters in magnitude due to the loss of wave energy.

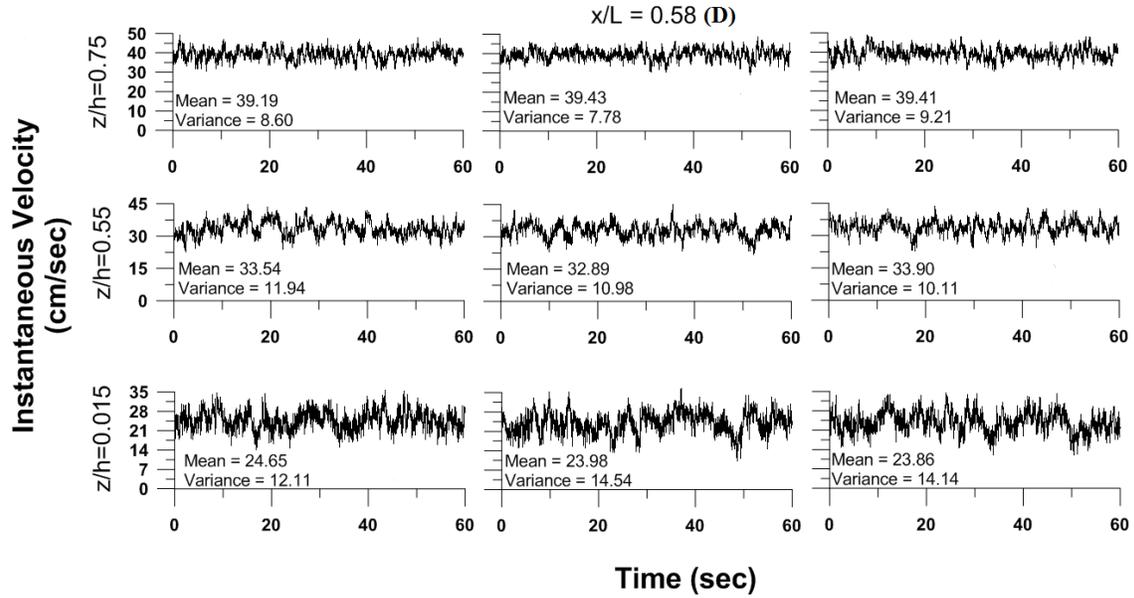

**Fig. 3** Time series plots of instantaneous velocity data for 60 sec each in wave-blocking condition at different levels (z/h) at location D, showing the equilibrium state.

**Table-1. Values of flow parameters at different discharges**

| Recorded discharge Q m³/s | Calculated discharge Qcal m³/s | Calculated discharge/Recorded | Maximum mean velocity (m/s) ($u_m$) | Reynolds number (Re) | Froude number ($Fr$) | Mean flow depth (m) ($h$) | Critical wave frequency (Hz) ($\omega_c$) | Wavelength (m) ($\lambda$) | Initial wave-amplitude (cm) ($h_w$) | Wave steepness (s) ($2\pi h_w/\lambda$) | $kh(=2\pi h/\lambda)$ | Wave nature |
|---|---|---|---|---|---|---|---|---|---|---|---|---|
| 0.0197 | 0.036 | 1.83 | 0.24 | 0.72x10⁵ | 0.14 | 0.30 | 1.4 | 0.29 | 0.70 | 0.15 | 6.47 (>π) | deep water |
| 0.0250 | 0.048 | 1.92 | 0.32 | 0.95x10⁵ | 0.18 | 0.30 | 1.31 | 0.36 | 0.80 | 0.14 | 5.21 (>π) | deep water |
| 0.0322 | 0.060 | 1.86 | 0.40 | 1.20x10⁵ | 0.23 | 0.30 | 1.3 | 0.45 | 1 | 0.14 | 4.18 (>π) | deep water |

## 3. Data processing and analysis

The collected velocity data were always affected by the Doppler noise associated with the measuring technique and thus noise has to be removed before the computation of turbulence parameters. The raw velocity data collected from all vertical locations were processed to remove noise using a phase space threshold de-spiking technique described by Goring and Nikora [31] and implemented in the win-ADV software Wahl [32]. The velocity data were analyzed systematically for all the locations. The effects of large noise were removed by minimizing the possible aliasing effect near the Nyquist frequency (herein $f_n$ = 20 Hz). About 2 to 4% of the raw data were excluded. Such excluded data signals were replaced by cubic polynomial interpolation technique.

In turbulent flow, the mean velocity is described with the aid of a Reynolds-like decomposition, where the periodicity of mean flow is accounted for. The instantaneous velocity components (u, v, w) in the Cartesian coordinate system (x, y, z) include a mean and a fluctuating component given by

$$u = \bar{u} + u'; \quad v = \bar{v} + v'; \quad w = \bar{w} + w' \tag{1}$$

where $\bar{u}$, $\bar{v}$, $\bar{w}$ = mean velocities along (x, y, z) directions; u′, v′, w′ = the fluctuations of velocity components u, v, w. Because the turbulent flows associated with eddies are random in motion, these are characterized by statistical concepts. The mean stream-wise, lateral and bottom-normal velocity components ($\bar{u}$, $\bar{v}$, $\bar{w}$) are defined as

$$\bar{u} = \frac{1}{n}\sum_{i=1}^{n} u_i, \quad \bar{v} = \frac{1}{n}\sum_{i=1}^{n} v_i \quad \text{and} \quad \bar{w} = \frac{1}{n}\sum_{i=1}^{n} w_i \tag{2}$$

where n is the number of observations (n = 7200) during full sample period 180 sec. The root-mean-squares of stream-wise and bottom-normal turbulent intensities ($\sigma_u$, $\sigma_w$) are defined as

$$\sigma_u = \sqrt{\overline{u'^2}} = \sqrt{\frac{1}{n}\sum_{i=1}^{n}(u_i - \bar{u})^2} \tag{3}$$

$$\sigma_w = \sqrt{\overline{w'^2}} = \sqrt{\frac{1}{n}\sum_{i=1}^{n}(w_i - \bar{w})^2} \tag{4}$$

Eqs (3) and (4) are denoted as the standard deviations of the 'random' stream-wise and bottom-normal velocity fluctuations. The larger standard deviation indicates a higher level of turbulence. The intensity of turbulence shows the degree of turbulence in the flow, which is defined by using the fluctuating component of velocity in all directions. The mean Reynolds shear stress component is defined as:

$$\tau_{xz} = -\rho \overline{u'w'} = -\frac{\rho}{n}\sum_{i=1}^{n}(u_i - \bar{u})(w_i - \bar{w}) \tag{5}$$

The normalized stream-wise ($I_u$) and bottom-normal ($I_w$) turbulent intensities and the normalized Reynolds shear stress ($\tau_{uw}$) are given by

$$I_u = \sigma_u / u_*, \quad I_w = \sigma_w / u_*, \quad \tau_{uw} = -\overline{u'w'} / u_*^2, \tag{6}$$

where $u_*$ is the shear velocity determined from the log-law.

To ensure the fully developed flow over the rigid flat surface, the collected velocity data are analyzed at all eight vertical locations (A to H) along the flume centerline, and the time-averaged stream-wise velocity are plotted in Fig. 4. It is observed that the vertical profiles of stream-wise mean velocity are constant along the flume centerline except the locations A and B in the upstream, indicating the development of turbulent boundary layer over the flat surface from A to H shown by a line. As the stream-wise velocity profiles along the flume at six locations (C to H) showed the fully developed flow, average of all six profiles were considered to plot the logarithmic stream-wise velocity and other turbulence parameters.

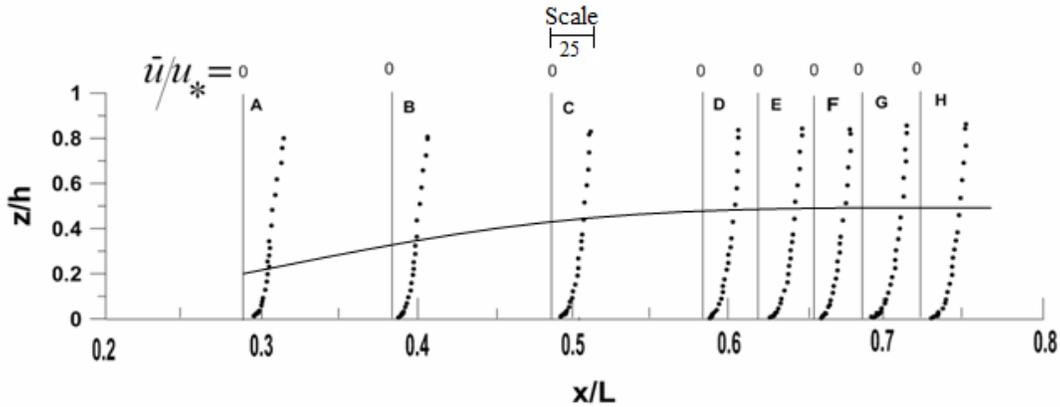

**Fig. 4** Stream-wise mean velocity profiles ($\bar{u}/u_*$) at eight horizontal locations (A, B…,H) along the flume centerline in flow condition and the line shows formation of turbulent boundary layer.

For hydraulically rough surface, the normalized stream-wise mean velocity ($\bar{u}/u_*$) was found to follow the log-law as

$$\bar{u}/u_* = 1/\kappa \ln(z/z_0), \tag{7}$$

where $\kappa$ is the von Karman constant (0.40) and $z_0$ is the equivalent bed roughness (equal to 0.0084 cm) with coefficient of regression $R^2 \approx 0.96$. Here, $u_*$ (= 1.598 cm/s) is the friction velocity and was determined from the log-law (Fig. 5a, b). The normalized lateral and bottom-normal mean velocity components ($\bar{v}/u_*, \bar{w}/u_*$) were approximately zero throughout the depth of the flow (Fig. 5a). The normalized turbulence intensity components ($I_u$ and $I_w$) and the Reynolds shear stress ($\tau_{uw}$) were plotted as a function of vertical distance (z/h) from the bed (Fig. 5c, d). It is observed from the figures that both the components of turbulence intensity do not vary significantly throughout the water depth (z/h), but stream-wise intensity shows a maximum value near the bed and then decreases slowly towards the water surface. Nezu and Rodi [33] proposed the following semi-analytical relationships for normalized turbulence intensities ($I_{u,w}$):

$$I_{u,w} = D_{u,w} \exp(-C_{u,w}\frac{z}{h}) \tag{8}$$

where $D_{u,w}$, $C_{u,w}$ are the dimensionless coefficients determined from the observed data of $I_{u,w}$ and are given by

$D_u = 2.1712$ and $C_u = 0.6786$ (9a)

$D_w = 0.7655$ and $C_w = 0.3322$ (9b)

The solid lines in Fig. 5c show the fitted values of intensities, which show in good agreement with Nezu and Rodi [33] and Nezu and Nakagawa [34]. However, these parameter values are not universal; they depend on the channel bed roughness. Fig. 5d shows the plot of normalized Reynolds shear stress ($\tau_{uw}$) against vertical distance (z/h). The maximum value of normalized shear stress ($\tau_{uw}$) occurs near the bed, and then decreases towards the water surface. The figure shows the normalized profiles of mean stream-wise, lateral and bottom-normal velocities, stream-wise and bottom-normal turbulence intensities and Reynolds shear stress of fully developed flow over the rigid flat surface.

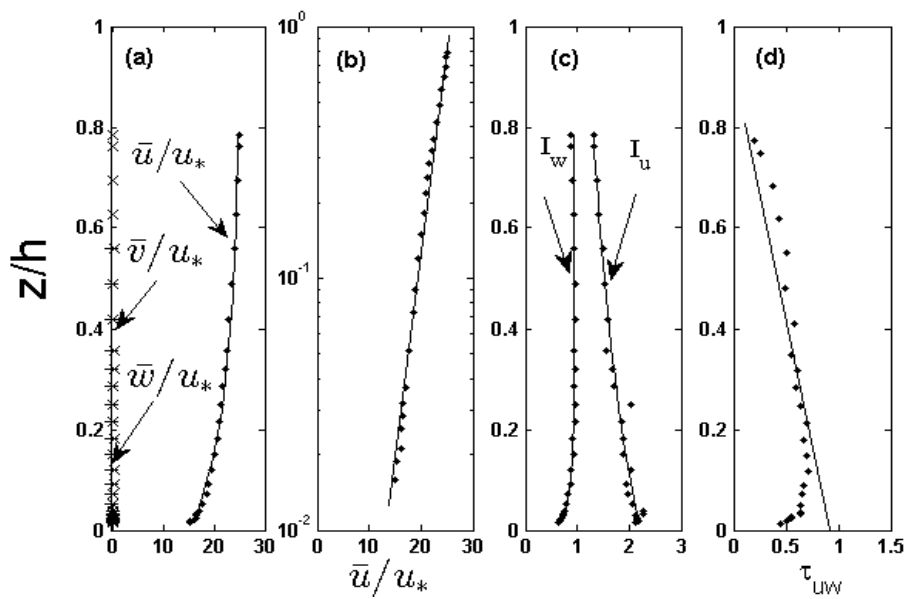

**Fig. 5** Normalized profiles: (a) symbol ● for stream-wise ($\bar{u}/u_*$); × for lateral ($\bar{v}/u_*$); and + for bottom-normal ($\bar{w}/u_*$) mean velocities, (b) Stream-wise velocity ($\bar{u}/u_*$) in log-scale, (c) Turbulence intensities ($I_u$ and $I_w$), (d) Reynolds shear stress ($\tau_{uw}$). Solid lines indicate fitted equation plot. These profiles are the average of all six profiles (C to H).

In order to verify the accuracy of individual time series velocity data of all three components, each of the velocity signals was first low pass filtered using Butterworth infinite impulse response (BIIR) with cut-off frequency 5 Hz to remove noise and to eliminate the possible aliasing effect. To minimize the noise in the higher frequency region of the intended spectra, each of the velocity signals was divided into ensembles of 1024 data points and hence 7 ensembles out of the total observed data were obtained. All of them were de-trended about their mean separately. The power spectra were estimated at 0.0195 Hz intervals from 0 to 20 Hz with 95% confidence limit using the power spectral density (PSD) algorithm available in the Matlab software package with Hamming window (Venditti and Bennett [35], Sarkar et al. [36]). Fig. 6 shows the plots of PSDs versus spectral frequency (*f*) of de-spiked velocity data of all three components in log-log scale at all six locations (C to H) at a mid-depth *z/h* = 0.53 over the flat surface along the flume. The slopes of the PSD follow the Kolmohorov -5/3 scaling-law within the inertial sub-range. Hence, it is confirmed the accuracy of the measurement of velocity samples at all three directions as well as the fully developed flow along the channel.

According to researchers (Yang et al. [37], Absi [38]), a two-dimensional flow in the central portion of the flume was achieved with width/depth ratio $\approx 2$. Since in the present study, the width/depth ratio was $\approx 2.1$ with the occurrence of maximum velocity at the flow depth $h = 24$ cm as dip-phenomenon, it is ascertained the two-dimensional flow at the central portion of the flume, and hence the effect of secondary currents due to side walls on the flow was negligible. The quantitative confirmation of lateral and bottom-normal mean velocity profiles throughout the depth is shown in Fig. 5a. Therefore, it was ensured that the flow was free from the secondary currents at the flume central line.

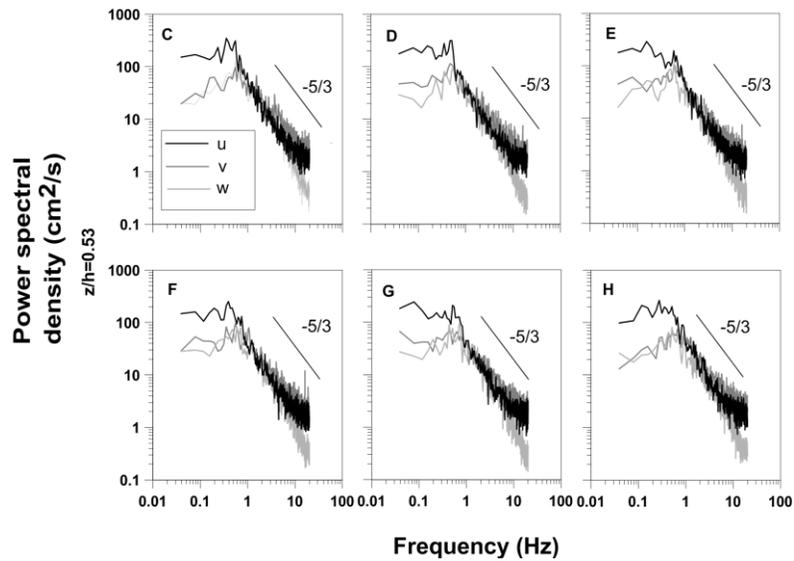

**Fig. 6:** Plots of power spectral density (PSD) against spectral frequency of all three velocity components at a level $z/h = 0.53$ along six locations (C, D,…, H) on the fully developed flow.

However, when waves are superimposed on a current, it is important to introduce the phase-averaged quantities because the instantaneous velocity is modified by the mean velocity, the time-dependent velocity and the fluctuating velocity due to wave component. Accordingly, the instantaneous velocity components *u* and *w* are usually take the form as (Nielsen [39]; Umeyama [9]):

$$u = \bar{u} + \tilde{u}(t) + u' \tag{10}$$

$$w = \bar{w} + \tilde{w}(t) + w' \tag{11}$$

where $\tilde{u}(t) = \langle u(t) \rangle - \bar{u}$ and $\tilde{w}(t) = \langle w(t) \rangle - \bar{w}$ with the over-bar denotes the mean velocity; tilde (~) denotes wave-induced velocity; prime denotes the fluctuating velocity; and the symbol $\langle \rangle$ denotes phase-averaged velocity. The wave-induced velocity $\tilde{u}(t)$, possesses a periodic nature, is obtained by subtracting the mean velocity $\bar{u}$ from the phase-averaged velocity $\langle u(t) \rangle$. The phase-averaged velocity $\langle u(t) \rangle$ is determined by computing the average over an ensemble of samples taken at a fixed phase in oscillation. The phase-averaged velocity is defined by (Umeyama [9], Singh et al. [13]):

$$\langle u \rangle = \left[ \frac{1}{N} \sum_{n'=0}^{N-1} u(t + n'T) \right] \tag{12}$$

where $n'$ = oscillation cycle number, $T$ is the wave period, $N$ is the total number of oscillation cycles. The phase-averaged velocity $\langle u(t) \rangle$ can be extracted from instantaneous velocity $u$ by cross-correlation with a sinusoidal wave in phase with oscillation. Here $N = 50$ sinusoidal waves-cycles are reproduced for velocity time series at a measurement point. However, this wave cycles are not observed in other cases, like wave-blocking region or further upstream. The normalized phase-averaged stream-wise and bottom-normal turbulence intensities are defined as:

$$\langle I_u \rangle = \frac{\langle \sigma_u \rangle}{u_*} = \frac{1}{N} \sum_{i=0}^{N} \sqrt{(u(t + i\Delta t) - \langle u(t + i\Delta t) \rangle)^2} / u_* \tag{13}$$

$$\langle I_w \rangle = \frac{\langle \sigma_w \rangle}{u_*} = \frac{1}{N} \sum_{i=0}^{N} \sqrt{(w(t + i\Delta t) - \langle w(t + i\Delta t) \rangle)^2} / u_* \tag{14}$$

where $N$ = total number of oscillation cycles; and the normalized phase-averaged Reynolds shear stress is defined as:

$$\langle \tau_{uw} \rangle = \frac{1}{N} \sum_{i=0}^{N} (u(t + i\Delta t) - \langle u(t + i\Delta t) \rangle)(w(t + i\Delta t) - \langle w(t + i\Delta t) \rangle) / u_*^2 \tag{15}$$

In the present study, calculations of turbulence parameters for wave-induced conditions were performed by phase-averaging of the velocity signal (Mattioli et al. [40], Umeyama [9, 10]).

Fig. 7 shows the raw velocity data against time at three different locations A, D. and H (namely, flow region, wave-blocking and wave-dominated regions) along the centreline at three different vertical levels ($z/h$ = 0.015, 0.183, and 0.77) for a pair of flow discharge ($Q$) and frequency ($\omega_c$) as ($0.0322\, m^3/s$, 1.30 Hz). The sinusoidal nature of combined wave-current flow showed prominence at the wave region H at the downstream, and this nature progressively decreased as approaching towards the bed as well as upstream of the flow. It is observed from Fig. 7 that the sinusoidal nature of wave diminishes close to the bottom surface ($z/h \leq 0.183$). Hence the phase-averaging was not feasible because of non-sinusoidal nature near the bottom boundary ($z/h \leq 0.183$) as well as away from the wave-dominated region along upstream. Therefore, in the present study the phase-averaged velocity was considered above the level $z/h \geq 0.183$. The wave propagation against the flow leads to a non-sinusoidal nature in the flow beyond the blocking region, indicating the loss of wave energy as approaching upstream. It focuses a regime known as 'wave-blocking', for which there is a stream-wise location where the wave propagation velocity vanishes, that means, the wave cannot penetrate over the blocking region. Therefore, the whole region of flow is divided into three distinct regions i.e. flow region, wave-blocking and wave region.

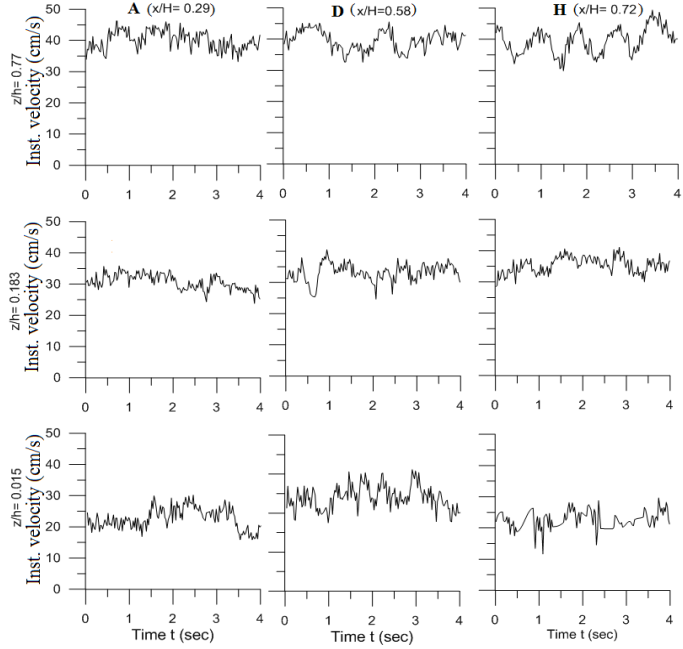

**Fig. 7** Stream-wise instantaneous velocity records for three different regions i.e. flow region (*x/H* = 0.29 at location A), wave-blocking (*x/H* = 0.58 at location D) and wave-dominated region (x/H = 0.72 at location H) at three different levels (z/h = 0.015, 0.183, and 0.770).

## 4. Dispersion relation for wave-blocking and verification with observed data

The present setup is completely adopted to study the specific feature of wave-current interaction, where the linear dispersion relation is primarily used for the low amplitude waves. Chawla and Kirby [16] showed that the blocking phenomena strongly depended on initial wave steepness. The waves are blocked perfectly in the opposing current when the wave steepness (*s* < 0.16) is small. In the present case, waves are not steep enough, and cannot produce wave breaking in the experimental condition (Table-1).

The linear dispersion relation for incompressible and small amplitude gravity wave on a fluid surface is given by

$$\omega^2 = gk \tanh(kH) \tag{16}$$

where $\omega$ is the frequency of wave, $H$ is the water depth, and $k$ is the wave vector. For a moving fluid medium, dispersion relation modifies as ([16, 19]):

$$\sigma^2 = (\omega - \vec{k}.\vec{U})^2 = gk \tanh(kH) \qquad (17)$$

where $\sigma$ is the wave frequency in moving frame, and $\vec{U}$ is the fluid velocity. Differentiating $\sigma$ with respect to $\vec{k}$, first two terms of Eq. (17) give

$$\vec{c}_{ga} - \vec{U} = \vec{c}_g \qquad (18)$$

where $\vec{c}_{ga} = \partial\omega/\partial\vec{k}$ is the group velocity vector in stationary frame and $\vec{c}_g = \partial\sigma/\partial\vec{k}$ is the group velocity vector in moving frame. The wave-blocking occurs at a point where the absolute group velocity vanishes, that is, at the blocking point $c_{ga} = 0$, Eq. (18) gives as (only in horizontal direction):

$$U + c_g = 0 \qquad (19)$$

Finally, differentiating $\sigma$ with respect to $k$, one can derive the group velocity $c_g$ from Eq. (17) as:

$$c_g = \frac{1}{2}(1 + \frac{2kh}{\sinh 2kh})\frac{\sigma}{k} \qquad (20)$$

Now using the equation (19), we verify our present experimental data for three different pairs of discharge and frequency $(Q, \omega_c)$ shown in Table-2, where $c_{ga} = \omega_c \lambda$, the group velocity in stationary frame (Table-1) and the fluid velocity U is assumed as the depth-averaged velocity ($U_{avg}$). It is observed from the Table-2 that the ratio of $U_{avg}/c_g$ is less than one for all three cases, indicating approximately the wave-blocking condition.

**Table-2: Verification of wave-blocking:**

| Recorded discharge Q ($m^3/s$) | Group velocity in stationary frame $c_{ga}$ (m/s) | Depth-averaged velocity $U_{avg}$ (m/s) | Group velocity in moving frame $c_g$ (m/s) | Ratio $U_{avg}/c_g$ |
|---|---|---|---|---|
| 0.0197 | 0.4060 | 0.1837 | 0.2223 | 0.83 |
| 0.0250 | 0.4716 | 0.2100 | 0.2616 | 0.80 |
| 0.0322 | 0.5850 | 0.2800 | 0.3050 | 0.92 |

However, several interesting and important features related to opposing wave-current interaction, i.e. non-linearity and frequency downshifting (Ma et al. [17] and Shugan et al. [18]) are not considered in the present set-up. The rigorous analysis requires for the Stokes non-linear third order dispersion law for the gravity wave, and the frequency downshifting becomes relevant mainly when wave breaking occurs for sufficiently steep waves. Shugan et al. [18] stated when a strong increase of wave steepness was observed within blocking region, leading to a wave breaking. The larger values of wave steepness *s* lead to wave-breaking. If the steepness is high enough, the main effect in wave-blocking area is wave breaking and frequency downshifting, which are not studied here. In the present experimental study the values of wave steepness lies below 0.16 (Table 1). So, the condition of blocking was fulfilled well in the present case.

## 5. Experimental results and discussions

### 5.1. Mean velocity components

The stream-wise mean velocity ($\bar{u}/u_*$) normalized by friction velocity $u_*$ are plotted against normalized vertical distance (*z/h*) in Fig. 8 along the flow at eight locations A to H with corresponding distances (*x/L*= 0.29, 0.38, 0.49, 0.58, 0.63, 0.66, 0.69 and 0.72) to illustrate the flow characteristics along the three different flow regimes: (flow, wave-blocking, and wave-dominated). Phase-averaged stream-wise velocity $\overline{\langle u \rangle}/u_*$ is plotted at G and H in the wave-dominated region. In the figure, the altered velocity profiles due to wave-blocking condition are compared with that of fully developed flow along the locations C to H.

It is observed that the mean velocity ($\bar{u}/u_*$) due to wave-blocking condition is less than that of fully developed flow (log-law) at all horizontal locations (C to G) except the extreme downstream location H where it merges with fully developed flow (log-law) apart from the near-free surface. Although the values of mean velocity in wave-blocking condition are less compared to fully developed flow, the vertical lines up to mid-depth (*z/h* = 0.5) apparently show that the mean velocity increases gradually from upstream to downstream, which is also confirmed from the Fig. 9, where all eight vertical velocity profiles due to wave blocking are plotted together. Kemp and Simons [5] and Umeyama [9, 10] conducted several experiments for waves opposing a current, but they did not focus their attention to the wave-blocking for which wave propagation velocity vanishes at certain stream-wise location. In combined wave-current flows the wave-blocking condition appears from a certain pair of flow discharge and frequency of opposing wave, which is rather different from that of Kemp and Simons [5] and Umeyama [9, 10].

Umeyama [9, 10] showed that the stream-wise mean velocity for waves against current was reduced throughout the depth for higher frequencies (WCA1 and WCA2) except near the free surface, when compared with logarithmic profile. The present mean velocity near the wave-blocking and wave-dominated regions shows a good agreement with that of Umeyama [9] for the cases WCA1 and WCA2 ($\omega$ = 1.11, 1.0 Hz) for waves against current. He also showed that the decrease of frequency (WCA2, WCA3, and WCA4) leads to increase of mean velocity near the free surface, when waves propagate against a current. This is probably because when waves propagated against current with low frequency; there is a loss of wave energy due to increase of wave attenuation, resulting increase in mean velocity near the water surface. The gradual transfer of wave energy to the opposing current reflects the change of mean velocity from downstream to upstream passing through wave-blocking location. Particularly, the magnitude of mean velocity due to wave-blocking significantly increases specially at mid-depth from upstream to downstream locations A to H.

It is also interesting to note that at the upstream locations A to C the velocity $\bar{u}/u_*$ due to blocking condition clearly indicates an inflection point at a level z/h ~ 0.4, then it disappears from the location D further downstream up to H, where the phase-averaged velocity $\overline{\langle u \rangle}/u_*$ decreases near the surface due to the dominated wave. It is observed from the figure that the mean stream-wise velocity $\bar{u}/u_*$ at all locations follows almost log-law with a shift from the fully developed flow. However, in combined wave-current flow, Kemp and Simons [5] and Umeyama [9] showed the Eulerian-mean velocity profile due to waves propagating against current started to deviate from the log-law in the region near the bed and increased towards free surface.

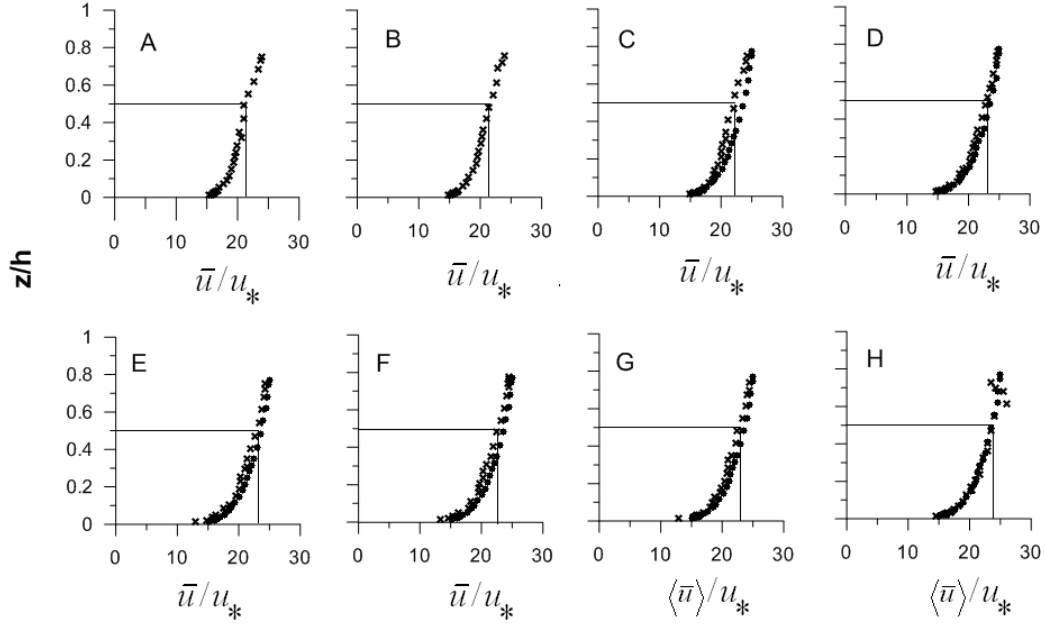

**Fig. 8**: Mean stream-wise velocity plots at eight selected locations along the flow. Symbol • represents for fully developed flow, and **x** (cross sign) represents for wave-blocking condition along the flow. Locations A, B, C indicate the flow region; locations D, E, F indicate the wave-blocking region; and locations G, H indicate the wave-dominated region. The vertical lines indicate the shifting of mean velocity at the mid-depth.

For clear visualization of velocity structures, the vertical profiles of stream-wise mean velocity from upstream to downstream at eight different locations (A to H) are also shown together in Fig. 9. It is observed that near-bed region up to the level $z = 3$cm, the stream-wise velocity is almost similar. The magnitude of mean velocity increases from upstream to downstream in the mid-depth, and subsequently it decreases near the bottom as well as little below the water surface to maintain the conservation of mass. The longitudinal velocity profiles along the flow seem to be strongly affected due to the opposing waves against current.

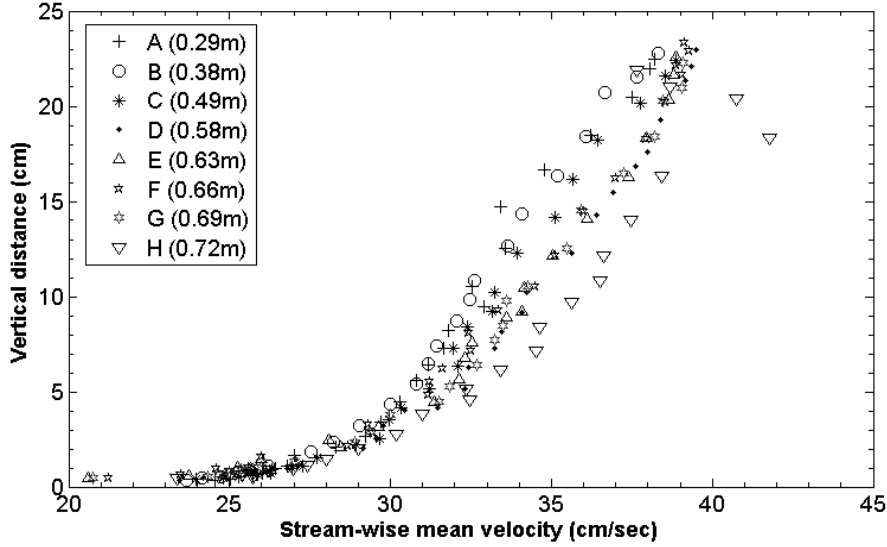

**Fig. 9**: Stream-wise velocity plots in eight different locations (A to H): each symbol represents respective vertical locations at different distances.

The profiles of normalized bottom-normal mean velocity along the flow are presented against $z/h$ in Fig. 10 for different locations A to H; and the phase-averaged velocity $\overline{\langle w \rangle}/u_*$ is plotted in locations G and H in the wave-dominated region. It is observed that the bottom-normal velocity is negative throughout the depth at all locations (A to H) due to opposing waves, and interesting to note that the bottom-normal velocity is directed downward due to wave-blocking condition. It is observed that the velocity $\overline{w}/u_*$ does not vary greatly up to the depth ~ 0.4 at the locations A-D, then it negatively increases up to the water surface; whereas at further downstream locations E-H, the velocity starts to increase linearly with negative value from the near-bed to the water surface. It is noticed that the magnitude of $\overline{w}/u_*$ decreases from the flow region to the blocking region, and again it increases in the wave region. The change in direction of $\overline{w}/u_*$ along downward depth indicates a circulation within the flow region due to wave-blocking, and consequently the stream-wise mean velocity ($\overline{u}/u_*$) increases more in core region along with the induced flow.

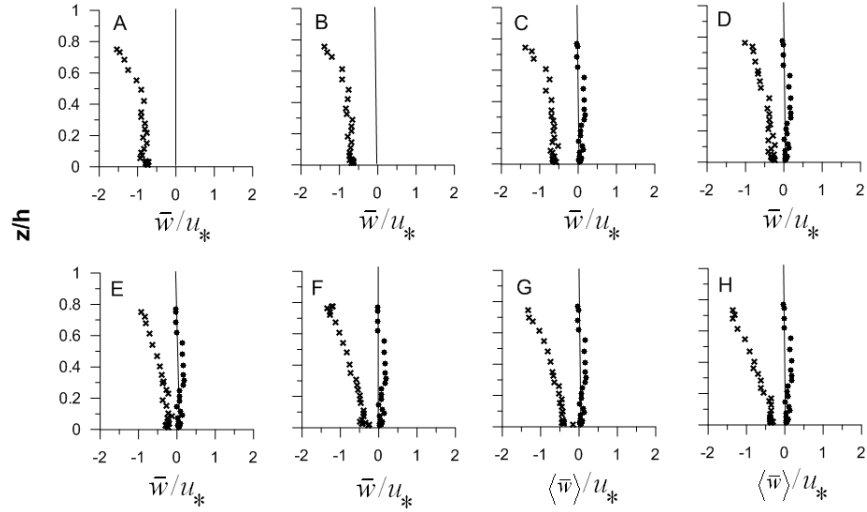

**Fig. 10**: Normalized bottom-normal mean velocity against *z/h* at eight locations (A to H). Symbol • represents for fully developed flow, and x represents for wave-blocking condition along the flow. Here locations A, B, C indicate the flow region; locations D, E, F indicate the wave-blocking; and locations G, H indicate the wave-dominated region.

*5.2. Turbulence intensities*

The normalized stream-wise and bottom-normal intensity ($I_u = \sigma_u/u_*, I_w = \sigma_w/u_*$) profiles are respectively plotted against vertical depth (z/h) in Figs. 11 and 12 at eight different locations along the flume:- at flow region (locations A, B, C), blocking region (locations D, E, F) and the wave region (locations G, H). The profiles of phase-averaged intensities ($\langle I_u \rangle = \langle \sigma_u \rangle/u_*, \langle I_w \rangle = \langle \sigma_w \rangle/u_*$) are plotted for the locations G and H of the wave-dominated region. In general, Fig. 11 reveals that the intensity $I_u$ increases at each location near the bed and reaches a maximum value and then decreases with increasing vertical distance from the bottom up to the level about ~0.15, and then no variation with the depth up to the location F; whereas at the wave-dominated region G and H, the value of $\langle I_u \rangle$ decreases gradually from the bed up to the level ~0.2 and no change up to the mid-depth ~0.5, and then again increases gradually with depth up to the water surface, which shows a concave nature in the profile. The present experimental results agree well with that of Kemp and Simons [5] and Umeyama [9] for the cases of WCA1 and WCA2, where the frequencies of waves are almost similar. Umeyama [9] reported that the superposition of the following and opposing waves over current, for both cases, $\langle I_u \rangle$ showed significant reduction from bed to mid-depth but no such effect near the surface, which also agrees well for the present case. Overall it is observed that the values of $I_u$ due to the wave-blocking at all locations are greater than that of the fully developed case, though the stream-wise mean velocity due to wave-blocking collapses with that of fully developed case at the location H.

It is observed from the Fig. 12 that the bottom-normal intensity $I_w$ increases all along the flow with respect to that of fully developed condition, but increment of $I_w$ is not that significant in the blocking region, whereas the increment of phase-averaged intensity $\langle I_w \rangle$ is high enough in the wave-dominated region at G and H, mostly near the surface. Increase of $\langle I_w \rangle$ is much prominent above z/h > 0.5. It is observed that the intensity $I_w$ at the locations C to F was almost constant from near-bed region to the water surface, whereas the phased-averaged $\langle I_w \rangle$ was constant up to the mid-depth, and then increased linearly to the water surface, which agrees well with the results of Umeyama [9] for case of wave against a current. The profiles of $I_w$ along the stream-wise locations show the similar trend as $I_u$ with smaller in magnitude except in the flow region A to C, where the intensity profiles show convex nature. It is noted from the figures that the overall trend of both the normalized intensity components is less at the blocking region compared to that of the flow and wave regions, which may be due to suppression of eddy movement due to blocking phenomenon. In fact, Kemp and Simons [5] and Umeyama [9, 10] analysed the experimental data for turbulence intensities due to waves opposing a current but they did not elucidate their work towards the wave-blocking condition generated from a certain pair of flow discharge and the frequency of opposing waves.

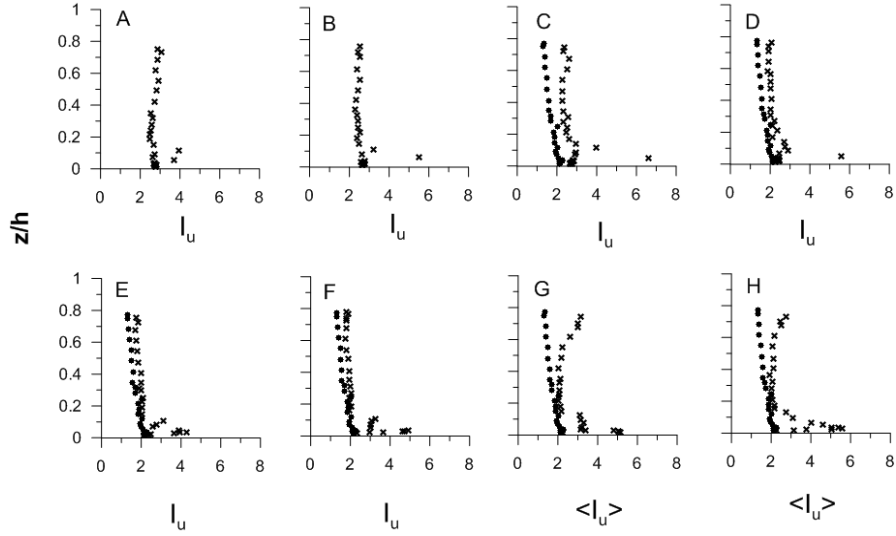

**Fig. 11**: Normalized stream-wise intensity against *z/h* at eight different locations: Symbol • represents for fully developed flow, and **x** (cross sign) represents for wave-blocking condition along the flow. Here locations A, B, C indicate the flow region; locations D, E, F indicate the wave-blocking region; and locations G, H indicate the wave-dominated region.

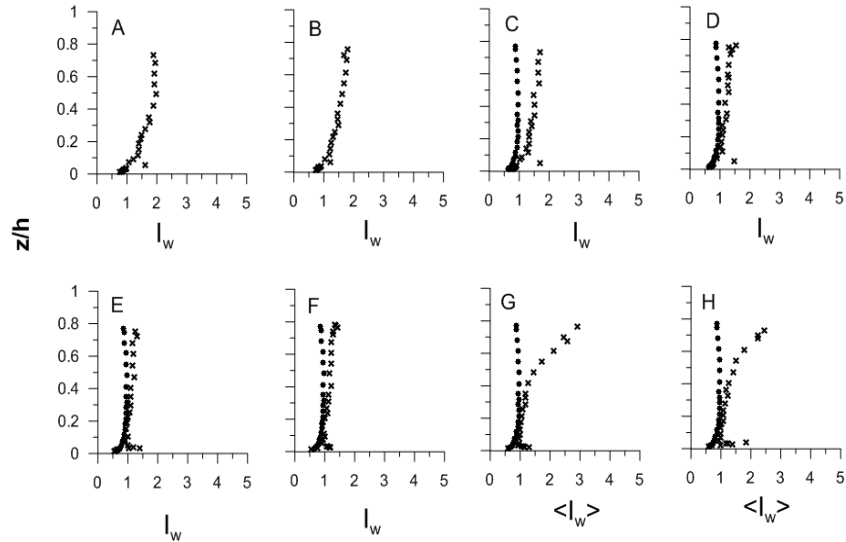

**Fig. 12**: Normalized bottom-normal intensity against *z/h* at eight different locations: Symbol • represents for fully developed flow, and **x** (cross sign) represents for wave-blocking condition along the flow. Here locations A, B, C indicate the flow region; locations D, E, F indicate the wave-blocking region; and locations G, H indicate the wave-dominated region.

*5.3. Reynolds shear stress*

The profiles of normalized Reynolds shear stress due to the wave-blocking are plotted against z/h in Fig. 13 along eight locations (A to H) including fully developed shear stress from C to H along the flow. It is observed from the figure that in the flow region (A, B, C) the shear stress ($\tau_{uw} = -\overline{u'w'}/u_*^2$) is almost same near the boundary up to z/h = 0.3, then it increases and reaches a maximum value at z/h = 0.7 and then again decreases, whereas for further downstream locations D to F, the qualitative nature of shear stress is almost same with the fully developed shear stress with higher magnitude at the mid-depth. But at the wave-dominated region (G and H), the normalized shear stress $\langle \tau_{uw} \rangle$ increases from the bottom, reaches a maximum value at the level z/h ≈ 0.35, then decreases and crosses that of fully developed flow at z/h = 0.6, indicating the lesser shear stress than that of fully developed condition near the free surface, which agrees well with that of Umeyama [9] for the case WCA1. The overall observations along the flow indicate that the normalized $\tau_{uw}$ is modulated differently in different regions due to the opposing waves on the current (Fig. 13). This phenomenon is possibly due to the fact that when the waves oppose the flow, the incidence wavelengths are seen to be decreased as approaching towards the blocking region, implying that there is loss of energy of the surface waves through the energy transfer due to mixing between the flow and the opposing wave, and hence the internal structure of turbulence is changed, consequently the Reynolds shear stress is also. However, the Reynolds shear stress for wave-blocking condition from upstream to downstream is positive throughout the depth. Umeyama [10] observed that the addition of waves in a current for both following and opposing reduced the turbulent shear stresses at all horizontal locations. A clear reduction in the phase-averaged shear stress $\langle \tau_{uw} \rangle$ was observed near the bed, which also has the similar trend in the present observation (Fig. 13). The distribution of shear stress can be expressed as

$$-\frac{\langle \overline{u'w'} \rangle}{u_*^2} = a'\left(\frac{z}{h}\right) + b', \tag{21}$$

which fits well with normalized shear stress profiles at the locations G to H with different coefficients $a'$ and $b'$; when $z = h$, shear stress is $a' + b'$, and when z = 0, that will be $b'$. Similarly, at locations E and F, normalized shear stress is computed and plotted with the experimental data.

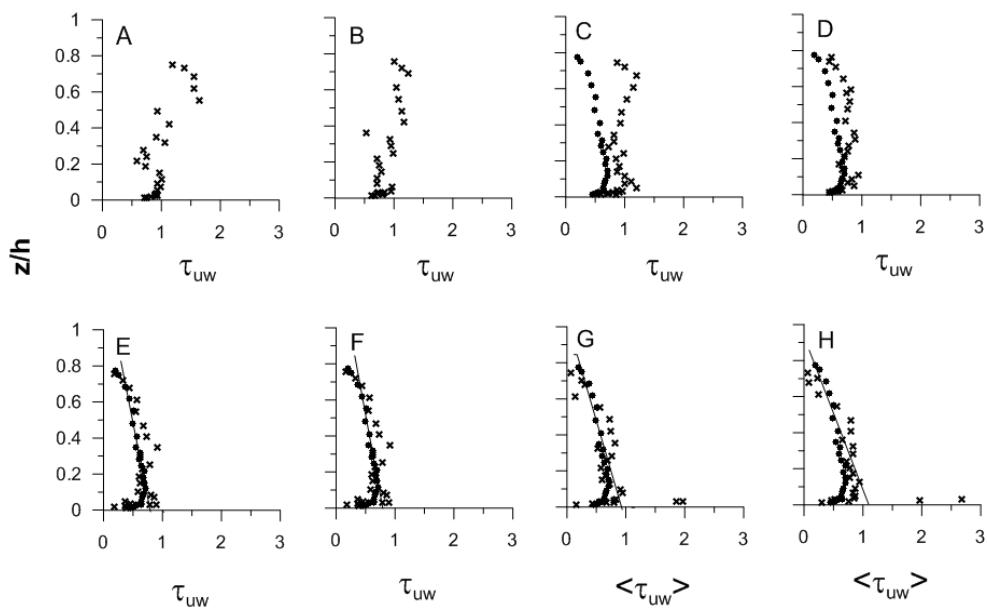

Fig. 13: Normalized Reynolds shear stress against *z/h* at eight locations along the flow: Symbol • represents for fully developed flow, and **x** (cross sign) represents for wave-blocking condition. Locations A, B, C indicate the flow region; locations D, E, F indicate the wave-blocking region; and locations G, H indicate the wave-dominated region Solid lines indicate fitting curve in wave-blocking condition using Eq. (21).

*5.4. Spectral and co-spectral analysis along the flow due to wave-blocking*

In order to understand the internal structure of turbulence due to the wave-blocking condition, the spectral analysis was used to detect noise in the velocity signals. In a similar manner as before, each of the velocity signals was divided into ensembles of 1024 data points, and hence 7 ensembles are obtained from the total observed data. Fig. 14 shows a sequence of power spectral density (PSD) plots of velocity signals against spectral frequency (*f*) in log-log scale for three selected vertical levels (z/h = 0.77, 0.183, 0.015) at eight locations A to H along the flow. The slope of the power spectra within the inertial sub-range was compared to the Kolmohorov -5/3 scaling-law. In Fig. 14, row represents three different water depths (*z/h*) and column represents eight locations (*x/L*):

locations (A, B, C) at flow region, locations (D, E, F) at wave-blocking and the locations (G, H) at wave-dominated region. The power spectra of stream-wise velocity signals suggest approximately a good fit to the slope -5/3 power-law having a moderately large inertial sub-range at heights z/h = 0.77 and 0.183. In wave-dominated region (x/$L$ = 0.72 at H), the figure shows that there is sharp peak in frequency 1Hz near the water surface. So, the value of PSD becomes maximum near about frequency 1 Hz in wave region, and from downstream to upstream the peak value diminishes gradually. It is clear that the effect of wave becomes smaller from wave-blocking region to flow region. At the near-bed region (z/h = 0.015), power spectra become flat, so the slopes are very small. It is obvious that there is some difference between the levels especially in the lowest level (z/h = 0.015). According to Corvaro et al.'s [41] observations, moving towards bottom from water surface, the PSD transferred to the high frequencies and energy attenuation became larger. He also observed that PSD was closer to -5/3 law near the surface and became flatter near the bed. Similarly, in the present case, the PSDs near the water surface approximately fitted with -5/3 law and become flatter near bed. It was also observed that the flatness was greater in wave region compared to the other regions. The reason may be that there is a disturbance near the bed. So, the flow is not isotropic in this location. The spectra of velocity would be identical at all levels if the flow was completely isotropic. It is noted that surface energy or power is maximum at the surface of wave-dominated region, and it reduces towards wave-blocking region and finally there is no frequency peak at the flow region at A. This trend shows in the whole vertical profiles in Fig. 14. So the surface wave energy diminished gradually from the wave region to the flow region. It was observed that PDSs showed a distinct behaviour for different levels, mostly near the blocking and wave-dominated regions, which might be related to the different turbulent energy levels.

To discuss the energy distribution elaborately, the co-spectral analysis is also examined for all eight locations. The analysis is performed on the joint time series of stream-wise and bottom-normal velocity components in a similar process shown in Fig. 15. In a one-dimensional velocity spectrum (e.g. u or w), the variance is proportional to the Reynolds normal stress (i.e. $\overline{u'}^2$ or $\overline{w'}^2$), whereas the covariance between two velocity components is proportional to the Reynolds shear stress ($\overline{u'w'}$). The co-spectral tends to have larger peaks than the corresponding stream-wise spectra (Venditti and Bauer [42]). Fig. 15 shows definite and dominant peak at flow region near the surface. Its values decrease near the bed. It is evident from Fig. 15 that the velocity co-spectra are flat and relatively undistinguished for $z/h = 0.015$, especially in wave regions (G and H) which suggest the absence of recurring oscillatory motions at dominant frequency zone (Venditti and Bauer [42]), which also supports the Corvaro et al. [41]'s analysis. At the mid-depth $z/h = 0.183$, the values of co-spectra become greater than that at the level $z/h = 0.77$ for both wave-blocking and wave regions (D to H).

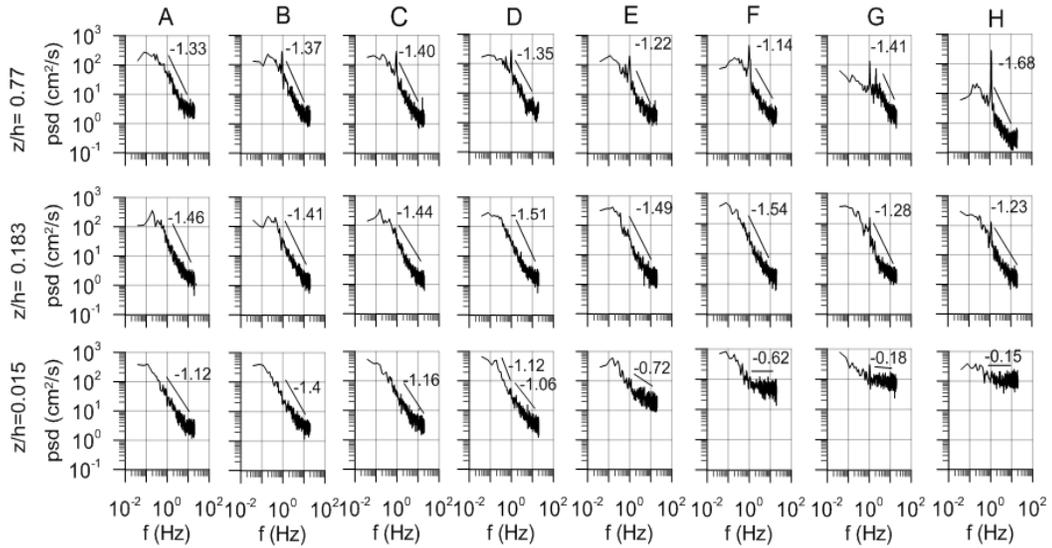

**Fig. 14**: PSD in eight different vertical locations (A to H) from upstream to downstream at wave-blocking condition. Row represents three different heights ($z/h = 0.015, 0.183, 0.77$) and column represents eight locations (A to H) along the flow.

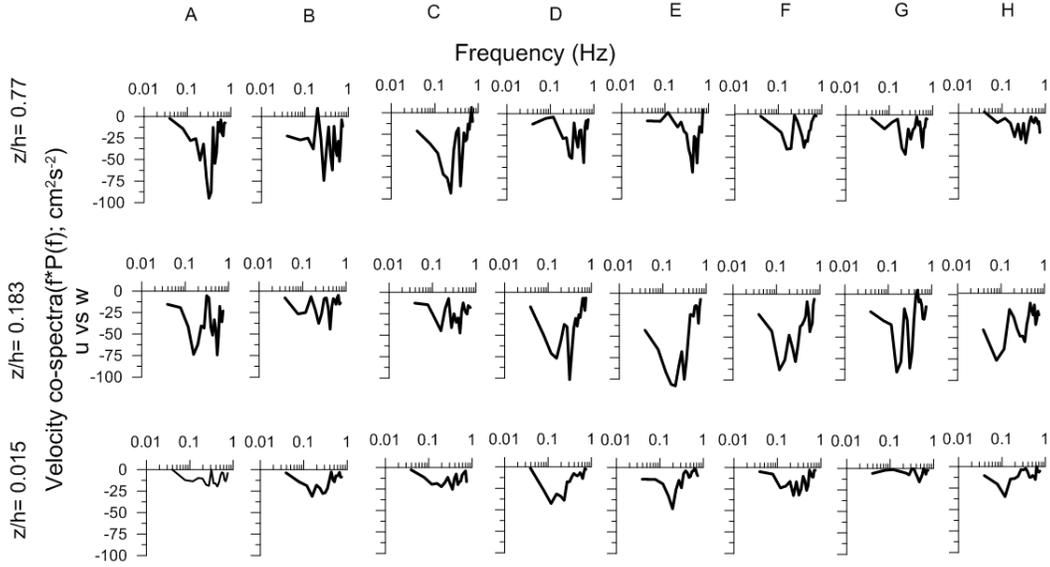

**Fig. 15**: Co-spectral plots in eight different locations (A to H) from upstream to downstream at wave-blocking condition. Row represents three different heights (z/h = 0.015, 0.183, 0.77) and column represents eight locations (A to H) along the flow.

*5.5. Eddy viscosity:*

The turbulent flow is often described by eddy viscosity as a local property of flow; as molecular viscosity which is a property of fluid. The closure models of the Reynolds shear stresses normally use an eddy viscosity hypothesis based on analogy between the molecular and turbulent motions. The turbulence eddies are thought of as lumps of fluid, like molecules, collide and exchange momentum. One of the most striking features of turbulence is the generation of turbulent eddies and self-similar behaviour over a range of scales (Batchelor and Townsend, [43]). These eddies play an important role in river turbulence, which affects the morpho-dynamics and sediment transport (Nikora and Goring, [44]). The corresponding mixing length due to the movement of turbulent eddies behaved like molecular mean free path derived from kinetic theory of gas (Markatos [45]). Since the Reynolds shear stress and the mean velocity gradient are known at different locations along the flow from upstream to downstream for fully developed flow and wave-blocking condition over the flat surface, the distribution of eddy viscosity $v_t^+$ with depth is computed according to Boussinesq's concept as:

$$v_t^+ = -\overline{u'w'} / (\partial \overline{u} / \partial z) \tag{22}$$

$$<v_t^+> = <-\overline{u'w'}> / (\partial \langle \overline{u} \rangle / \partial z) \tag{23}$$

for flow condition and phase-averaged condition respectively (Hussain and Reynolds [46], Nezu and Nakagawa [34]). The normalized eddy viscosity for flow and wave-blocking regions (locations A to F) is written from the log-law as:

$$v_t = v_t^+ / u_* h = \kappa \frac{z}{h}\left(1 - \frac{z}{h}\right), \tag{24}$$

and for wave-dominated region (locations G and H), the phase-averaged normalized eddy viscosity can be written as $<v_t> = <v_t^+> / u_* h$. The profiles of eddy viscosity are plotted against z/h in Fig. 16. Solid line indicates the observed data for eddy viscosity in a fully developed flow (current-alone basic flow) and the dashed line indicates the fitted curve passing through the observed data of eddy viscosity for wave-blocking condition. In a steady flow over the flat surface using logarithmic velocity distribution, the eddy viscosity is parabolic and approximately linear near the bottom as suggested by Nezu and Rodi [33] using experimental data. It is also observed from the figure that values of eddy viscosity due to the wave-blocking along the flow from upstream to downstream (locations A to H) show scattered with a parabolic nature throughout the depth. It is interesting to note that the normalized eddy viscosity due to wave-blocking condition increases away from the boundary (z/h > 0.3) along the flow from C to E with respect to the eddy viscosity ($v_t$) for fully developed flow over a flat surface, then decreases and collapses at the locations of wave dominated region, except the location H, where the eddy viscosity $\langle v_t \rangle$ decreases sharply near the surface (z/h > 0.5). Along the flow from upstream to downstream, two different behaviors of eddy viscosity appear due to wave-blocking condition- one from the region A to E, and other from F to H in the wave-dominated region. The occurrence of two different behaviors of eddy viscosity along the flow is probably due to the modulated shear stress in three different regions: flow in the upstream, wave-blocking at a location and opposing wave in the downstream regions. Kemp and Simons [5] stated that the values of eddy viscosity became more scattered due to the superposition of waves over current. A similar feature is also observed in the present case in the wave-dominated region (Fig. 16). The present results agree also with that of Grant and Madsen [2] and You et al.[47] model for combined wave-current flow over rough bed. However, in the region 0.1< z/h <0.4 the eddy viscosity is constant and thereafter decreases.

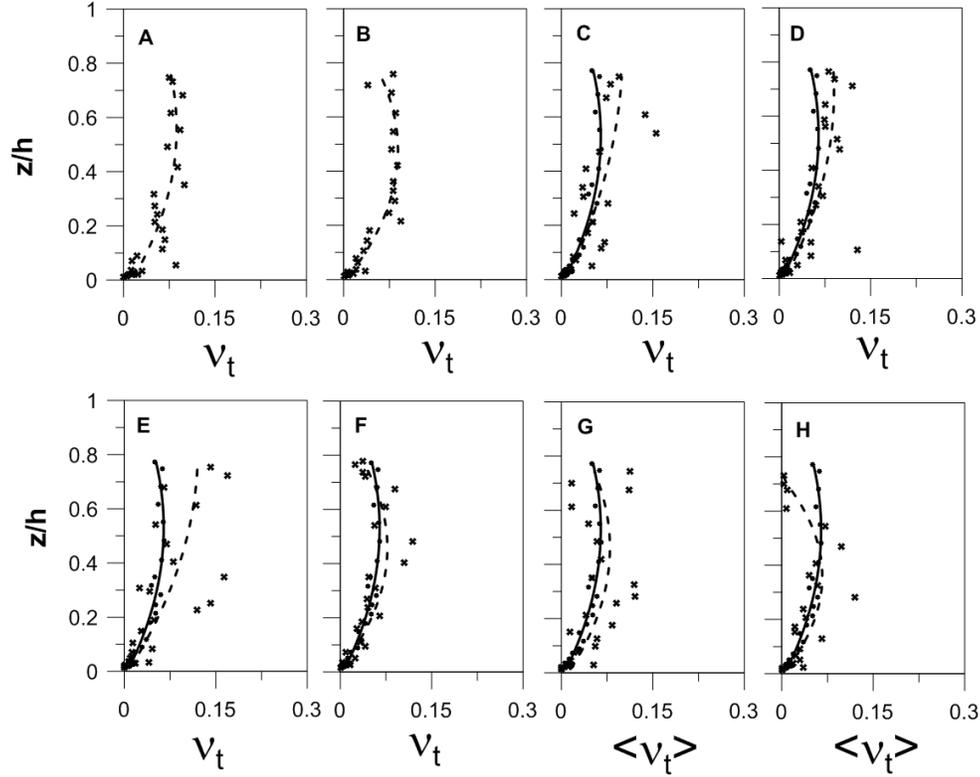

**Fig. 16**: Normalized eddy viscosity against *z/h* at eight locations along the flow: Symbol • represents eddy viscosity in fully developed flow, and **x** (cross sign) represents eddy viscosity during wave-blocking. Here locations A, B, C indicate the flow region; locations D, E, F for the wave-blocking region; and locations G, H indicate the wave-dominated region. Solid line indicates the fitted curve for eddy viscosity for fully developed flow, and the dashed line indicates the fitted curve in blocking condition.

*5.6. Turbulence kinetic energy (TKE) fluxes*

The stream-wise and bottom-normal turbulence kinetic energy (TKE) fluxes are calculated as

$$f_{ku} = \frac{0.5(\overline{u'u'u'}+\overline{u'v'v'}+\overline{u'w'w'})}{u_m^3} \qquad (25)$$

$$f_{kw} = \frac{0.5(\overline{w'u'u'} + \overline{w'v'v'} + \overline{w'w'w'})}{u_m^3} \tag{26}$$

for flow and wave-blocking regions at locations A to F, and

$$<f_{ku}> = \frac{0.5(\overline{<u'u'u'>} + \overline{<u'><v'v'>} + \overline{<u'><w'w'>})}{u_m^3} \tag{27}$$

$$<f_{kw}> = \frac{0.5(\overline{<w'><u'u'>} + \overline{<w'><v'v'>} + \overline{<w'w'w'>})}{u_m^3} \tag{28}$$

for wave-dominated region at G and H locations (Raupach [48], Maity and Mazumder [49]). The vertical profiles of normalized TKE fluxes are plotted against vertical height $z/h$ in Fig. 17 for stream-wise flux $f_{ku}$, $<f_{ku}>$; and in Fig. 18 for bottom-normal flux $f_{kw}$, $<f_{kw}>$ for wave-blocking condition at eight different locations (A to H) from upstream to downstream along with that of fully developed flow (current-alone) over the flat surface for C to D. Here both the fluxes of TKE due to the wave-blocking are compared with that of fully developed condition, where the stream-wise flux $f_{ku}$ is negative and the bottom-normal flux $f_{kw}$ is positive throughout the depth. The stream-wise flux $f_{ku}$ for wave-blocking case varies along the flow with changing sign throughout the depth for all locations A to H in an oscillatory nature. At the flow region A, B, C, the values of $f_{kw}$ are negative around the mid-depth; at the locations D, E, F, G and H in the wave-blocking and wave-dominated regions, the flux $f_{kw}$ is almost zero except near boundary for all locations and near the surface for the wave-dominated region at G and H. At downstream locations F to H both stream-wise and vertical TKE fluxes at near-bottom boundary show scattered positive and negative values. Agelinchaab and Tachie [50] showed that the flux $f_{kw}$ was positive near the flat surface boundary, whereas Balachandar and Bhuiyan [51] showed that the flux $f_{ku}$ was negative values for the smooth surface, which agree with the present observations. The negative and positive values of the flux $f_{ku}$ lead to the transport of energy in the backward and forward directions respectively; and related to the ejection–sweep character of the Reynolds shear stress. This is probably because the kinetic energy fluctuation is retained by extraction of energy from the mean flow. While negative and positive values of $f_{kw}$ indicate respectively the transport of energy in the forward and backward directions. It is observed from the figures that the vertical distributions of $f_{ku}$ shown in Fig. 17 are qualitatively similar to those of $f_{kw}$ in Fig. 18 with higher magnitude.

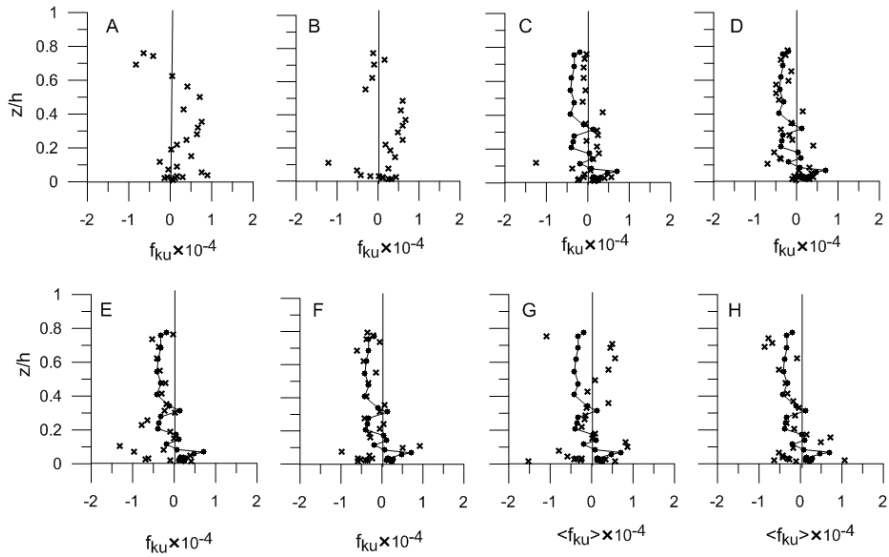

**Fig. 17**: Normalized TKE flux in *u*-direction against *z/h* along the flow at eight different locations: symbol • with solid line represents fully developed flow, and **x** (cross sign) represents wave-blocking condition. Here locations A, B, C indicate the flow region; locations D, E, F indicate the wave-blocking region; and locations G, H indicate the wave-dominated region.

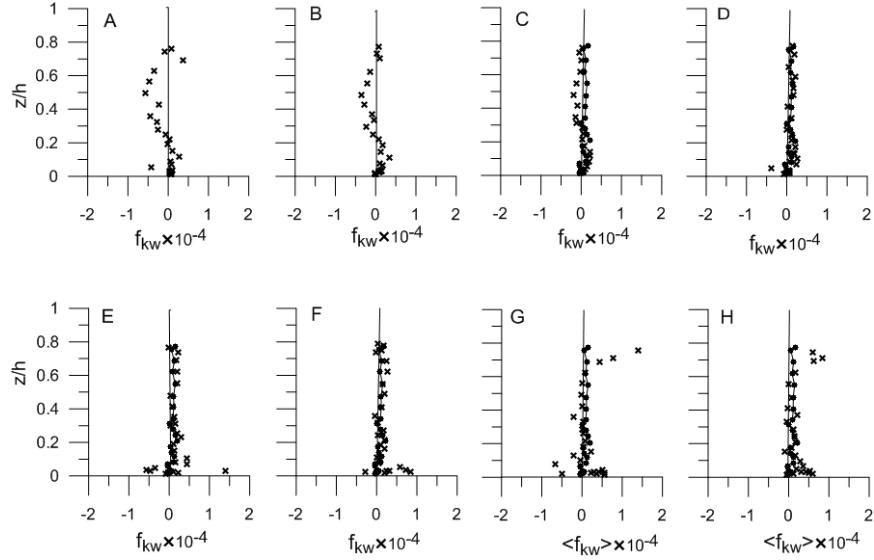

Fig. 18: Normalized TKE flux in *w*-direction against *z/h* along the flow at eight different locations: symbol • with solid line represents fully developed flow, and **x** (cross sign) represents wave-blocking condition. Here locations A, B, C indicate the flow region; locations D, E, F indicate the wave-blocking region; and locations G, H indicate the wave-dominated region.

## *5.7. Quadrant analysis due to wave-blocking condition*

The quadrant analysis of Reynolds shear stress is frequently used to quantify the existence of coherent structures in the flow and to itemise their contributions to the total shear stress [52, 53, 54, 32]. The quadrants are usually referred to as the following names: (a) Quadrant-1 (Q1), outward interactions ($i = 1$: $u' > 0, w' > 0$), (b) Quadrant-2 (Q2), ejections ($i = 2$: $u' < 0, w' > 0$), (c) Quadrant-3 (Q3), inward interactions ($i = 3$: $u' < 0, w' < 0$), and (d) Quadrant-4 (Q4), sweeps ($i = 4$: $u' > 0, w' < 0$). In this study, analysis is performed to highlight directly the contributions of turbulent event evolutions to the total shear stress for wave-blocking condition. At any point in a turbulent flow, the contribution of the total Reynolds shear stress from quadrant *i*, excluding the region $\mathcal{H}$, is defined as,

$$[u'w']_{i,\mathcal{H}} = \lim_{n\to\infty} \frac{1}{n} \int_0^n u'(t)w'(t) I_{i,\mathcal{H}}[u',w'] dt \tag{29}$$

where $n$ is the total number of observations and the square bracket denotes a conditional average; and the indicator function $I_{i,\mathcal{H}}$ is defined as

$$I_{i,\mathcal{H}}(u',w') = \begin{cases} 1, & \text{if } (u', w') \text{ is in the i}^{\text{th}} \text{quadrant and if } |u'w'| \geq \mathcal{H}|\overline{u'w'}| \\ 0, & \text{otherwise} \end{cases} \quad (30)$$

In the wave-dominated region, the phase-averaging is given by

$$[<u'w'>]_{i,\mathcal{H}} = \lim_{n\to\infty} \frac{1}{n} \int_0^n \langle u'(t)\rangle\langle w'(t)\rangle I_{i,\mathcal{H}}[\langle u'\rangle,\langle w'\rangle]dt \quad (31)$$

and the indicator function $I_{i,\mathcal{H}}(<u'>,<w'>)$ for phase averaging is

$$I_{i,\mathcal{H}}(\langle u'w'\rangle) = \begin{cases} 1, & \text{if } (\langle u'\rangle, \langle w'\rangle) \text{ is in the i}^{\text{th}}\text{quadrant and if } |\langle u'\rangle\langle w'\rangle| \geq \mathcal{H}|\overline{\langle u'\rangle\langle w'\rangle}| \\ 0, & \text{otherwise} \end{cases}$$

(32)

Here, $\mathcal{H}$ is the threshold parameter in the Reynolds shear stress signals by which one can extract the values of $<u'w'>$ from the whole set of data, which are greater than $\mathcal{H}$ times $|\overline{\langle u'w'\rangle}|$ value. The stress fraction by ith quadrant is defined as [49]:

$$S_{i,\mathcal{H}} = \frac{[u'w']_{i,\mathcal{H}}}{[\overline{u'w'}]}, \quad (33)$$

and for wave-dominated region using phase-averaging, the stress fraction $\langle S_{i,\mathcal{H}}\rangle$ is as

$$\langle S_{i,\mathcal{H}}\rangle = \frac{(<\langle u'\rangle\langle w'\rangle>)_{i,\mathcal{H}}}{[\overline{\langle u'\rangle\langle w'\rangle}]} \quad (34)$$

which gives the Reynolds shear stress fraction associated with each of the turbulent events. By definition for $\mathcal{H} = 0$, $S_{1,0} + S_{2,0} + S_{3,0} + S_{4,0} = 1$ for both flow and wave-blocking regions, and $\langle S_{1,0}\rangle + \langle S_{2,0}\rangle + \langle S_{3,0}\rangle + \langle S_{4,0}\rangle = 1$ for wave-dominated region.

Stress fraction $|S_{i,\mathcal{H}}|$ for all three different regions (locations A, D and H) at each quadrant are plotted against threshold parameter $\mathcal{H}$ in Fig. 19a for the height $z/h = 0.76$; in Fig. 19b for $z/h = 0.35$; and in Fig. 19c for $z/h = 0.142$, where the locations A, D, H are denoted by different symbols in the figure. It is observed that the contribution of ejections and sweeps to the total shear stress are much higher than that of inward and outward interactions for all threshold parameter $\mathcal{H}$ values except near the surface at z/h = 0.76. The contributions due to wave effect are much higher in all four quadrants for all $\mathcal{H}$ except at the level *z/h* = 0.35, where the contributions due to flow are much higher in all four quadrants. In quadrants Q2 and Q4, it is observed that as the wave propagates from downstream to upstream, the contributions gradually decreases from the wave region (location H) to the flow region (location A) near the surface at height z/h = 0.76. It is interesting to note that the higher values of stress fraction sustain in the wave-dominated region for all quadrants, whereas contributions are almost equal for the flow (location A) and wave-blocking (location D) regions at the levels *z/h* = 0.142. But at the level z/h = 0.35, contributions of all turbulent events to the total shear stress are greater for flow region (location A) than that of both wave-blocking and wave-dominated regions. The contributions of ejections and sweeps to the total shear stress are much higher than that of inward and outward interactions at the levels *z/h* = 0.35, 0.142, whereas the contribution are almost similar for all quadrants at the level *z/h* = 0.76. For example, at the level *z/h* = 0.76 for $\mathcal{H}$ = 5, the contributions from the quadrants are $S_{1,5}$ = -0.2076, $S_{2,5}$ = 0.7279, $S_{3,5}$ = -0.2723, $S_{4,5}$ = 0.6106, the sum of them is 0.8586, which is 86 % of the average shear stress $\tau$ for the flow region at the location A; the contributions from the quadrants are $S_{1,5}$ = -1.111, $S_{2,5}$ = 1.590, $S_{3,5}$ = -1.129, $S_{4,5}$ = 1.563, the sum of them is 0.913, which is 91 % of the average shear stress $\tau$ for the wave-blocking region at D; and the contributions from the quadrants are $<S_{1,5}>$ = -7.907, $<S_{2,5}>$ = 9.080, $<S_{3,5}>$ = -8.488, $<S_{4,5}>$ = 8.305, the sum of them is 0.99, which is 99 % of the average shear stress $\tau$ for the wave-dominated region at location H. Similarly, at the levels z/h = 0.35, and 0.142 for $\mathcal{H}$ = 5, 10 and 15, the contributions from all quadrants to the total shear stress are determined for all three regions at locations A, D and H. Table-3 shows the percentage of average shear stress $\tau$ for all three different regions A (flow), D (wave-blocking) and H (wave-dominated) at three different levels *z/h* = 0.76, 0.35, and 0.142 for three threshold parameter $\mathcal{H}$ = 5, 10, and 15. It is interesting to note that at the surface level z/h = 0.76 for any fixed threshold parameter $\mathcal{H}$, contribution in percentage of average shear stress increases from the flow region to the wave-dominated region along downstream, whereas at the other levels $z/h = 0.35$, and $0.142$, percentage contribution of average shear stress shows irregular behavior from the flow region to the wave-dominated region for different threshold parameters $\mathcal{H}$. This irregular behavior in percentage contributions of average shear stress near the bottom boundary may be due to mixing generated from the waves propagating against the flow. It is observed from the Fig. 9 that the stream-wise mean velocity increases from upstream to downstream in the mid-depth, and subsequently it decreases near the bottom as well as water surface from downstream to upstream to maintain the conservation of mass. The longitudinal velocity profiles along the flow seem to be strongly affected due to the opposing waves.

For clear visualization of the turbulent bursting events along the flow over three different regions (flow, wave-blocking and wave-dominated) due to wave-blocking condition, the regions of all four quadrant events (Q1, Q2, Q3 and Q4) are shown in Figs. 20(a-c) in red colour scale for three values of threshold parameters $\mathcal{H} = 0$, 5 and 10 respectively. The figures clearly show that the contributions of all four quadrant events to the total shear stress are dominant for the wave-dominated region compared to that of other two regions for all three values of $\mathcal{H}$. Interesting to note that in the wave region both Q2 and Q4 are more prominent than Q1 and Q3 for all $\mathcal{H}$ values. Particularly, for $\mathcal{H} = 0$, contributions of ejection (Q2) and sweep (Q4) at all three regions along the flow near the surface are dominant than that of outward (Q1) and inward (Q3 interactions. As the value of H increases, these contributions decrease in all three regions, especially in the flow and wave-blocking regions, but in the wave-dominated region all four events are dominant implying the higher shear stresses are generated mainly in the wave-dominated region. At the interface of two regions (wave-dominated and wave-blocking), the interaction of ejection and sweeps occurs, which may play an important role in the kolk-boils phenomena (Mao [55] , Ojha and Mazumder [56]). It has been observed that some sediment could be ejected from the sediment bed into suspension due to kolk-boils formation. It seems that inside the pathway of the kolk-boil, the fluid pressure is low enough to form a negative pressure gradient and pick up sediment particles into suspension.

**Table -3. Percentage (%) of average shear stress τ**

| Percentage (%) of average shear stress τ in conditional statistics for three different hole numbers at different locations (A, D, H) | | | | | | | | | |
|---|---|---|---|---|---|---|---|---|---|
| Water depth (z/h) | Flow region (A) (%) | | | Wave-blocking region (D) (%) | | | Wave-dominated region (H) (%) | | |
| | $\mathcal{H}=5$ | $\mathcal{H}=10$ | $\mathcal{H}=15$ | $\mathcal{H}=5$ | $\mathcal{H}=10$ | $\mathcal{H}=15$ | $\mathcal{H}=5$ | $\mathcal{H}=10$ | $\mathcal{H}=15$ |
| (a) 0.760 | 85.86 | 54.38 | 26.78 | 91.30 | 75.96 | 59.61 | 99.00 | 101.00 | 110.9 |
| (b) 0.350 | 88.08 | 65.97 | 44.00 | 68.04 | 27.89 | 8.76 | 68.04 | 12.24 | 8.76 |
| (c) 0.142 | 63.80 | 21.91 | 5.50 | 77.00 | 24.47 | 5.50 | 70.37 | 32.73 | 11.40 |

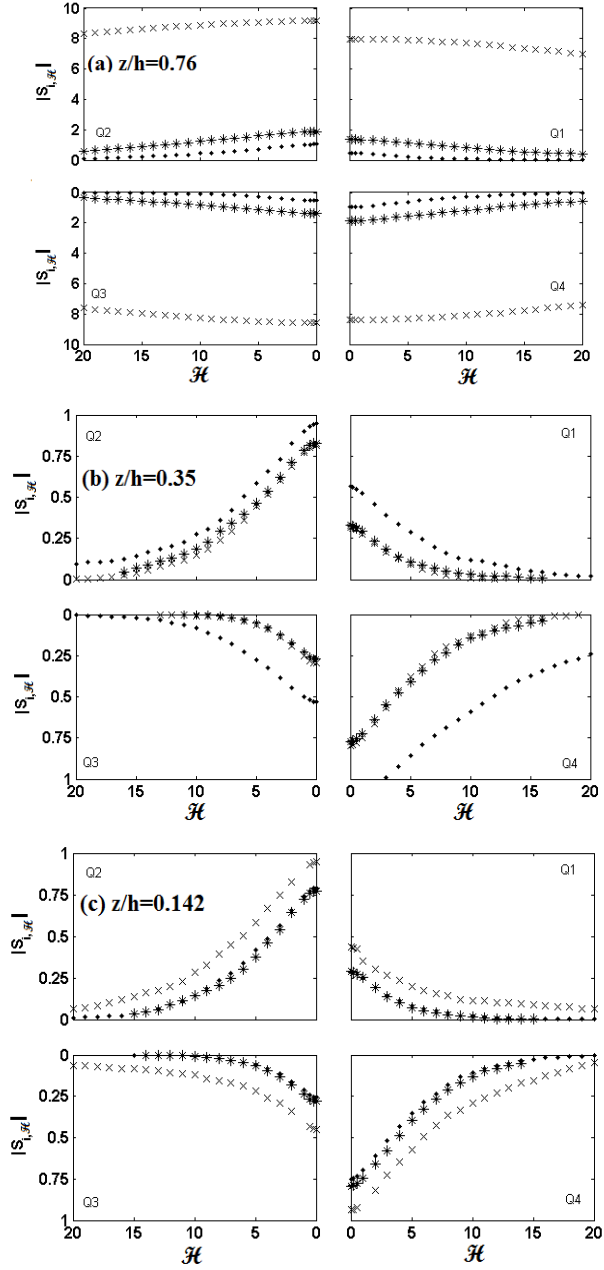

**Fig. 19**: Stress fraction $|S_{i,\mathcal{H}}|$ for each quadrant against $\mathcal{H}$: (a) vertical depth z/h = 0.76, (b) z/h = 0.35 and (c) z/h = 0.142. Symbol • denotes values in flow region at location A; * denotes values in blocking region at location D; and × denotes values in wave-dominated region at location H.

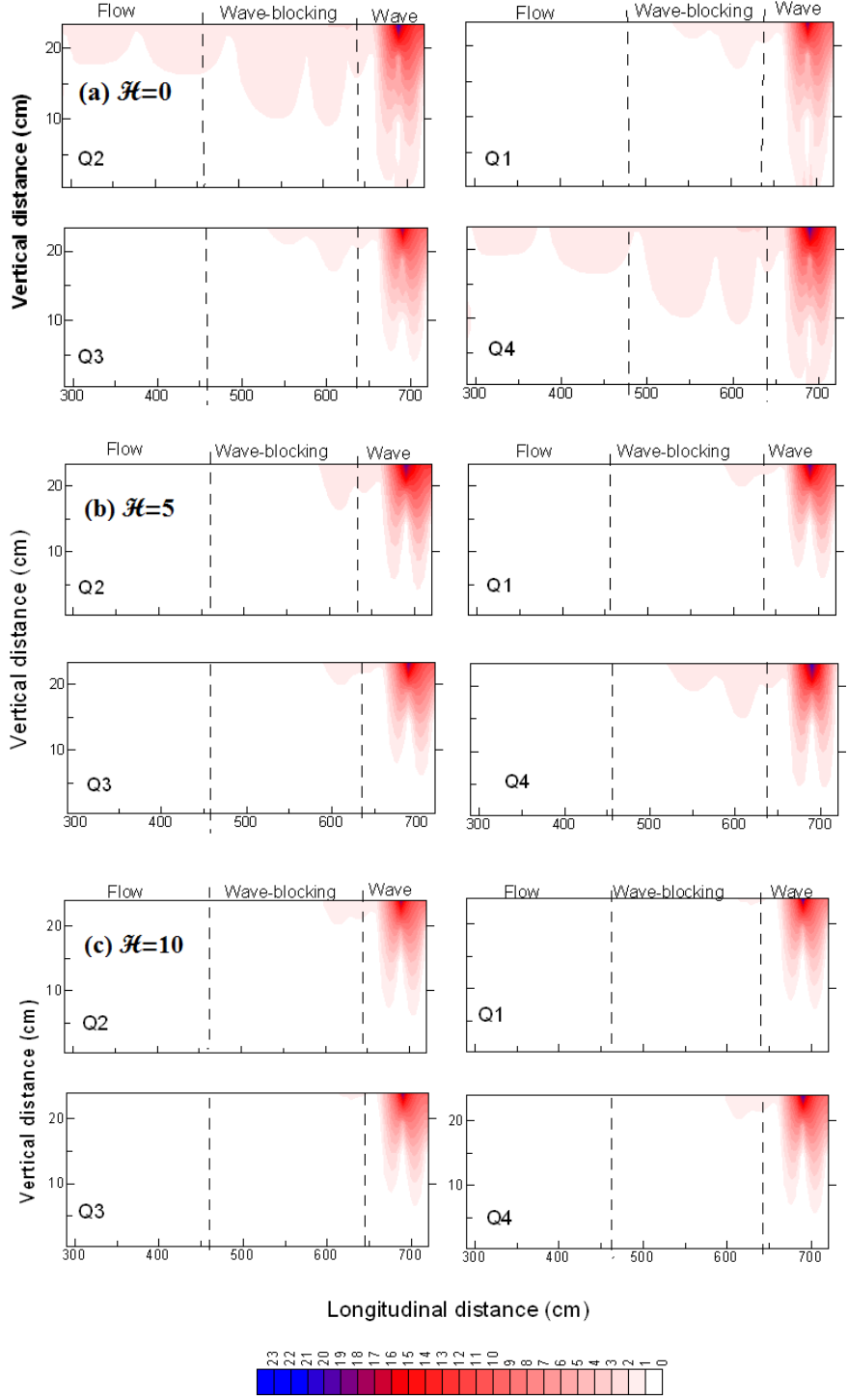

**Fig. 20**: Contour plots of stress fraction $|S_{i,\mathcal{H}}|$ for each quadrant in the xz-plane: (a) for $\mathcal{H} = 0$, (b) for $\mathcal{H} = 5$ and (c) for $\mathcal{H} = 10$.

## 6. Summary and conclusions

A laboratory study has been carried out for 'wave-blocking' phenomenon generated from the counter-current propagating waves. It is observed that the propagation of waves against the flow is blocked at a location where the counter-current reaches the group velocity of the wave. In fact, there exists a wave-blocking in a stream-wise location from a pair of flow discharge and frequency of counter propagating waves. A non-uniform flow develops throughout the flume due to counter propagating waves. The whole flow along the flume can be segmented into three regions, such as the flow region at the upstream, wave-blocking region at middle, and the wave-dominated region at the downstream. Here there is an impressed gravity wave against the flow and to the best of our knowledge there is not much literature both in the form of experimental results or theoretical analysis. Our aim is to explore the turbulent flow characteristics in the wave-blocking condition, which was not rigorously studied earlier in the flow with three different segmented regions.

Due to the generation of wave-blocking, a non-uniform distribution of mean flow was observed along the flume with three segmented regions. The magnitude of mean velocity increases from upstream to downstream in the mid-depth, and subsequently it decreases both at the near-bottom and water surface along the flume to maintain the conservation of mass. The mean velocity due to waves propagating against current is reduced towards upstream with respect to the fully developed flow. The transfer of wave energy to the opposing current reflects the sharing of mean velocity from downstream to upstream passing through wave-blocking.

Overall it is observed that the stream-wise ($I_u$) and bottom-normal ($I_w$) turbulence intensities due to wave-blocking case at all the horizontal locations are greater than that of the fully developed flow, but increment of $I_w$ is not so significant in the blocking region, whereas the increment of phase-averaged intensity $\langle I_w \rangle$ is high enough in the wave-dominated region, mostly near the surface. The stream-wise intensity ($I_u$) in each location reaches a maximum near the bed and then decreases with increasing vertical distance from the bottom. The variations of normal stresses $I_u$ and $I_w$ are more prominent near the bed and water surface to the wave region (G and H) compared to the other regions due to the effects of surface waves.

The decay in the energy level of the surface waves was explained from the velocity spectra analysis. Power spectral density (psd) of collected velocity data were determined at different locations from upstream to downstream during wave-blocking conditions to find the dominant frequency and the corresponding strength of different kind of vortices. It is observed that the power spectra of stream-wise velocity signals lead to a good fit to the slope -5/3 power law having a moderately large inertial sub range at the near-surface and mid-depth along horizontal direction, whereas interesting to that at the near-bed region, the slopes are very small, especially in the wave-dominated region.

Due to wave-blocking condition, TKE fluxes are compared with that of fully developed condition, where the stream-wise flux $f_{ku}$ is negative and the bottom-normal flux $f_{kw}$ is almost zero throughout the depth. Also it is interesting to note that the stream-wise flux $f_{ku}$ varies throughout the depth for all the locations A to H in an oscillatory nature. The negative and positive values of the flux $f_{ku}$ lead to the transport of energy in the backward and forward directions respectively; and related to the ejection–sweep character of the Reynolds shear stress. While negative and positive values of $f_{kw}$ indicate respectively the transport of energy in the forward and backward directions. It is also observed that the vertical distributions of $f_{ku}$ are qualitatively similar to that of $f_{kw}$ in higher magnitude.

It is also observed that values of eddy viscosity due to the wave-blocking condition along the flow from upstream to downstream show scattered with a parabolic nature throughout the depth. The values of eddy viscosity in the wave-blocking condition become larger from upstream to downstream than the fully developed flow.

The quadrant threshold technique is used to differentiate the three distinct regions. It is clearly observed that the contribution of ejections and sweeps to the shear stress are much higher than the contribution of inward and outward interactions. The effect of wave is clearly seen in four quadrants for $\mathcal{H} = 0$ and decays with increasing $\mathcal{H}$. The average shear stress τ shares 100 % near the surface in wave-dominated region for wave-blocking condition. Conditional statistics shows that there is some effect of waves near the bed in wave-blocking condition as the values of average shear stress τ at wave-dominated region share maximum percentage than flow and wave-blocking regions.

Here the experimental set-up was concentrated only to study a specific feature of wave-current interaction, where the linear dispersion relation was primarily used for the low amplitude waves for wave-blocking condition. However, several important features related to counter-current propagating waves, like non-linear dispersion and frequency downshifting [17, 18] were not considered in the present set-up. In fact, the Stokes non-linear third order dispersion law for the gravity wave and the frequency downshifting analysis become relevant mainly when wave breaking occurs for sufficiently steep waves. If the wave steepness is high enough, the main effect in wave-blocking area is wave breaking and frequency downshifting. Therefore, a detailed analysis is required considering the non-linearity and frequency downshifting effects to the counter-current propagating waves for high wave steepness; this will be studied separately to come out with generalized results.

This experimental study was directly performed to study the turbulence properties due to combined wave-current flows, especially when the waves propagated against the current with wave-blocking condition; though such a study has the potential to be useful to the researchers who study the turbulence behavior in natural environment.

**Acknowledgements**

It is indeed a pleasure to thank Silke Weinfurtner for suggestions regarding the experimental setup. We must acknowledge Professor Pengzhi Lin, Editor, and two anonymous Reviewers for their constructive comments and suggestions including some relevant references for the improvement of the paper, and Professor J. T. Kirby for providing us some of his papers on wave-blocking phenomena. We are grateful to Prof. Koustuv Debnath, Kausik Sarkar and Santosh Kumar Singh for helpful suggestions and discussions in data processing.

## References

[1] J.D.A. Van Hoften, S. Karaki, Interaction of waves and a turbulent current. Proc. of the 15th Int. Conf. on Coast. Eng. (New York) (ASCE) (1976) 404-422.

[2] W.D. Grant, O.S. Madsen, Combine wave and current interaction with rough bottom. J. Geophys. Res. 84 (4). (1979) 1797–1808.

[3] I. Brevik, B. Aas, Flume experiment on waves and currents: I. Rippled bed. Coastal Eng. 3. (1980) 149–177.

[4] P.H. Kemp, R.R. Simons, The interaction of waves and a turbulent current: waves propagating with the current. J. Fluid Mech. 116. (1982) 227–250.

[5] P.H. Kemp, R.R. Simons, The interaction of waves and a turbulent current: waves propagating against the current. J. Fluid Mech. 130. (1983) 73–89.

[6] G. Klopman, Vertical structure of the flow due to waves and current: laser-Doppler flow measurements for waves following or opposing a current Repository Hydraul. Eng. Report Part II Delft Hydraulic, Delft (1994).

[7] P.P. Mathisen, O.S. Madsen, Wave and currents over a fixed rippled bed: 1. Bottom roughness experienced by waves in the presence and absence of current. J. Geophys. Res. 101(C7). (1996a) 16533–16542.

[8] P.P. Mathisen, O.S. Madsen, Wave and currents over a fixed rippled bed: 2. Bottom and apparent roughness experienced by currents in the presence of waves. J. Geophys. Res. 101 (C7). (1996b) 16543–16550.

[9] M. Umeyama, Reynolds stresses and velocity distributions in a wave-current coexisting environment. J. Waterway Port Coastal Ocean Eng. 131(5). (2005) 203–212.

[10] M. Umeyama, Changes in turbulent flow structure under combined wave-current motions. J. Waterway, Port, Coastal, Ocean Eng, 135(5). (2009) 213-227.

[11] B.S. Mazumder, S.P. Ojha, Turbulence statistics of flow due to wave–current interaction. Flow Meas. Instrum. 18. (2007) 129–138.

[12] S.P. Ojha, B.S. Mazumder, Turbulence characteristics of flow over a series of 2-D bed forms in the presence of surface waves. *J*. Geophys. Res. 115, (2010) 1–15.

[13] S.K. Singh, K. Debnath, B.S. Mazumder, Turbulence statistics of wave-current flows over a submerged cube. J. Waterway, Port, Coastal, Ocean Eng. DOI:10.1061/(ASCE) WW.1943-5460.0000329, 04015027., (2015) 1-20.

[14] I.C. Van Rijn, M.W.C. Nieuwjaar, T.H. Van der Kaaij, E. Nap, H.F.A. Van Kampen, Transport of fine sands by current waves. J. Waterway, Port, Coastal, Ocean Eng. 119(2), (1993) 123-143.

[15] P. Nielson, Z. J. You, Eulerian-mean velocities under non-breaking waves on horizontal bottoms. Proc. Int. Conf. on Coastal Engineering, ASCE, New York, (1996) 4066-4078.

[16] A. Chawla, J. T. Kirby, Monochromatic and random wave breaking at blocking points, *J. Geophys. Res.*, 107 (2002), C7, 10.1029/2001JC001042.

[17] Y. Ma, G. Dong, M. Perlin, X. Ma, G. Wang, J. Xu, Laboratory observations of wave evolution, modulation and blocking due to spatially varying opposing currents. J. Fluid Mech., 661, (2010) 108-129.

[18] I. V. Shugan, H. H. Hwung, R. Y. Yang, An analytical model of the evolution of a Stokes wave and its two Benjamin-Feir sidebands on non-uniform unidirectional current. Nonlinear Processes in Geophysics. 22(3), (2015) 313-324. DOI: 10.5194/npg-22-313-2015.

[19] M. C. Haller, H. T. Özkan-Haller, Wave breaking and rip current circulation. Proc. of 28th Inter. Conf. on Coastal Engineering, Cardiff, UK, (2002) 705 – 717.

[20] M. Soltanpour, F. Samsami, T. Shibayama, S. Yamao, Study of irregular wave-current-mud interaction. Coastal Engineering Proceedings, (2014) DOI:10.9753/ ice.v34. waves.27.

[21] J.S. Ribberink, Time-averaged sediment transport phenomena in combined wave-current flows. Delft Hydraulics, *Part-II*, Report H-1889.(1995) 11.

[22] B.S. Mazumder, R.N. Ray, D.C. Dalal, Size distributions of suspended particles in open-channel flow over sediment beds. Environmetrics 16 (2005) 149–165.

[23] K. Sarkar, C. Chakraborty, B.S. Mazumder, Variations of bed elevations due to turbulence around submerged cylinder in sand beds. Environ Fluid Mech. 16 (2016) 659–693, DOI 10.1007/s10652-016-9449-0

[24] L.P. Euve, F. Michel, R. Parentani, G. Rousseaux, Wave blocking and partial transmission in subcritical flows over an obstacle. Phys. Rev. D, 91 (2015) 024020.

[25] SonTek Inc. ADV principles of operation, technical document, San Diego (2001).

[26] A. Lohrmann, R. Cabrera, N.C. Kraus, Acoustic-Doppler velocimeter (ADV) for laboratory use, in Fundamental and Advancements in Hydraulic Measurements and Experimentation, edited by C. A. Pugh, ASCE, (1994) 351-365.

[27] J. Bendat, A. Piersol, Random Data. Third ed. (2000) Wiley, New York.

[28] K. Barman, K. Debnath, B.S. Mazumder, Turbulence between two inline hemispherical obstacles under wave-current interactions. Adv. Water Resour. 88, (2016) 32–52.

[29] P. Maissa, G. Rousseaux, Y. Stepanyants, Recent results on the problem of wave-current interaction including water depth, surface tension/amplitude and vorticity effects. Coastal Dynamics (2013) 1137-1146.

[30] R.G. Dean, R.A. Dalrymple, Water wave mechanics for engineers and scientists. World-Scientific, Singapore, (2000).

[31] D.G. Goring, V.I. Nikora, Despiking acoustic Doppler velocimeter data. J. Hydraul. Eng. 128 (1) (2002) 117–126.

[32] T.L. Wahl, Analyzing ADV data using WinADV. In: Hotchkiss, R.H., Glade, M. (Eds.), Building Partnerships. ASCE, Reston (2000) (sect. 75, chap. 3).

[33] I. Nezu, W. Rodi, Open channel flow measurements with a laser Doppler anemometer. J. Hydraul. Eng., 112(5) (1986) 335–355, doi:10.1061/(ASCE)0733-9429(1986) 112:5 (335).

[34] I. Nezu, H. Nakagawa, Turbulence in Open-Channel Flows. A.A. Balkema, CRC Press, Rotterdam (1993).

[35] J.G. Venditti, S.J. Bennett, Spectral analysis of turbulent flow and suspended sediment transport over fixed dunes. J. Geophys. Res. 105 (C9) (2000) 22035–22047.

[36] K. Sarkar, B.S. Mazumder, Turbulent flow over the trough region formed by a pair of forward-facing bedform shapes. Europ. J. Mechanics B/Fluids. 46 (2014) 126-143.

[37] S.Q. Yang, S.K. Tan, S.Y. Lim, Velocity distribution and dip-phenomenon in smooth uniform open cannel flows. J. Hydraul. Eng. 130(12) (2004) 1179-1186.

[38] R. Absi, An ordinary differential equation for velocity distribution and dip-phenomenon in open channel flows. J. Hydraul. Res. 49 (2011) 82–89.

[39] P. Nielsen, Coastal bottom boundary layers and sediment transport. Advance Series on Ocean Eng 4. 1992, World Scientific Publ. Co., Singapore.

[40] M. Mattioli, A. Mancinelli, M. Brocchini, Experimental investigation of the wave-induced flow around a surface-touching cylinder. J. Fluids and Struct. 37 (2013) 62-87.

[41] S. Corvaro, E. Seta, A. Mancinelli, M. Brocchini, Flow dynamics in porous medium. Coastal Engineering, 91 (2014) 280-298.

[42] J.G. Venditti, B.O. Bauer, Turbulent flow over a dune. Green River Colorado. Earth Surf. Processes Landf. 30 (2005) 289–304.

[43] G.K. Batchelor, A.A. Townsend, The Nature of Turbulent Motion at Large Wave-Numbers, Proc. Roy. Soc. London. Series A, Mathematical and Physical Sciences, (1949) DOI: 10.1098/rspa.1949.0136.

[44] V.I. Nikora, D.G. Goring, Eddy convection velocity and Taylor's hypothesis of 'frozen' turbulence in a rough-bed open-channel flow, Journal of Hydro-science and Hydraulic Engineering, JSCE (2000) 18 (2) 75-91.

[45] N.C. Markatos, The mathematical modeling of turbulent flows. Applied Mathematical Modelling, 10 (1986) 190-220.

[46] A.K.M.F. Hussain, W.C. Reynolds, Measurements in fully developed turbulent channel flow. J Fluids Eng, ASME 97 (1975) 568-580.

[47] Z. J. You, D.L. Wilkinson, P. Nielsen, Velocity distribution of wave and current in combined flow. Coastal Engineering, 15 (2001) 525-543. http://dx.doi.org/10.1016/0378-3839 91 90026-D.

[48] M.R. Raupach, Conditional statistics of Reynolds stress in rough wall and smooth-wall turbulent boundary layers. J. Fluid Mechanics 108 (1981) 363–382.

[49] H. Maity, B.S. Mazumder, Experimental investigation of the impacts of coherent flow structures upon turbulence properties in regions of crescentic scour. Earth Surf Process Landform, 39 (2014) 995–1013. Doi:10.1002/esp.3496.

[50] M. Agelinchaab, M. F. Tachie, Open channel turbulent flow over hemispherical ribs J. Heat Fluid Flow 27 (2006) 1010–27.

[51] R. Balachandar, F.Bhuiyan, Higher-order moments of velocity fluctuations in an open-channel flow with large bottom roughness J. Hydraul. Eng. 133 (2007) 77–87.

[52] E.R. Corino, R.S. Brodkey, A visual investigation of the wall region in turbulent flow. J. Fluid Mech. 37(1) (1969) 1–30.

[53] S.S. Lu, W.W. Willmarth, Measurements of the structure of the Reynolds stress in a turbulent boundary layer. J. Fluid Mech. 60(3) (1973) 481-511.

[54] H. Nakagawa, I. Nezu, Prediction of the contributions to the Reynolds stress from bursting events in open channel flows. J. Fluid Mechanics 80(1) (1977), 99–128.

[55] Y. Mao, The effects of turbulent burstings on the sediment movement in suspension. Int. J. Sediment Res. 18(2) (2003) 148–157.

[56] S.P. Ojha, B.S. Mazumder, Turbulence characteristics of flow region over a series of 2-D dune shaped structures. Advances in Water Resources 31 (2008) 561–576.